\documentclass [useAMS,usenatbib]{mn2e}

\usepackage{amssymb,amsmath}
\usepackage{epsfig}
\usepackage{color} 

\begin{document}

\title[Magnetic Connection Model for Launching Jets ]{Magnetic Connection Model for Launching Relativistic Jets from a Kerr Black Hole}

\author[Ioana Du\c{t}an]{Ioana Du\c{t}an\thanks{Member of the International Max Planck Research School (IMPRS) for Astronomy and Astrophysics at the Universities of Bonn and Cologne; E-mail:
idutan@mpifr-bonn.mpg.de}\\Max-Planck-Institut f\"{u}r Radioastronomie, Auf dem H\"{u}gel 69, 53121 Bonn, Germany}

\date{2010 January 22}

\maketitle

\begin{abstract}
We present a model for launching relativistic jets in active galactic nuclei (AGN) from an accreting Kerr black hole (BH) as an effect of the rotation of the space-time, where the gravitational energy of the accretion disc inside the ergosphere, which can be increased by the BH rotational energy transferred to the disc via closed magnetic field lines that connect the BH to the disc (BH-disc magnetic connection), is converted into jet energy. The main role of the BH-disc magnetic connection is to provide the source of energy for the jets when the mass accretion rate is very low. We assume that the jets are launched from the disc inside the BH ergosphere, where the rotational effects of the space-time become much stronger, being further accelerated by magnetic processes. Inside the ergosphere, we consider a split topology of the magnetic field, where parts of the disc connect to the BH and other parts to the jets via magnetic field lines. The rotation of the space-time channels a fraction of the disc energy (i.e., the gravitational energy of the disc plus the rotational energy of the BH which is deposited into the disc by magnetic connection) into a population of particles that escape from the disc surfaces, carrying away mass, energy and angular momentum in the form of jets, allowing the remaining disc gas to accrete. 
In this picture, the BH can undergo recurring episodes of its activity with: (i) a first phase when the accretion power dominates and (ii) a second phase when the BH spin-down power dominates. In the limit of the spin-down power regime, the model proposed here can be regarded as a variant of the Blandford-Znajek mechanism, where the BH rotational energy is transferred to the disc inside the ergosphere and then used to drive the jets. As a result, the jets driven from a disc inside the BH ergosphere can have a relatively strong power for low mass accretion rates. We use general-relativistic conservation laws to calculate the mass flow rate into the jets, the launching power of the jets and the angular momentum transported by the jets for BHs with a spin parameter $a_* \geqslant 0.95$. As far as the BH is concerned, it can (i) spin up by accreting matter and (ii) spin down due to the magnetic counter-acting torque on the BH. We found that a stationary state of the BH ($a_* = $ const) can be reached if the mass accretion rate is larger than $\dot{m} \sim 0.001$. The maximum value of the BH spin parameter depends on $\dot{m}$ being less but close to 0.9982 (Thorne's model). In addition, the maximum AGN lifetime can be much longer than $\sim 10^{7}$ yr when using the BH spin-down power. 
\end{abstract}

\begin{keywords}
accretion, accretion discs -- black hole physics -- magnetic
fields -- galaxies:jets 
\end{keywords}

\section{Introduction}

Relativistic jets are highly collimated plasma outflows present in extragalactic radio sources, which are associated with many AGN. If the matter accreted by a BH (at the centre of an AGN) has enough angular momentum compared to that of a particle moving in a circular orbit around and near the BH, an accretion disc can be formed. The launching power of the AGN jet can generally be provided by the accretion disc, by the BH rotation, or both. Moreover, as the jet is launched, the BH can evolve towards a stationary state with a spin parameter whose maximum value is less but close to one ($a_* \lesssim 1$, where $-1\leq a_*\leq +1$). One can consider the launching power of the jet to be a fraction of the disc power. A number of questions come to mind: Is this fraction generally valid for astrophysical jets from BHs with the same mass and spin? Can the disc manage to launch the jet by itself as the BH accretes at low rates? How does the magnetic field get involved? Can the BH take over and support the disc to launch the jet as the mass accretion rate goes down? How does the BH spin evolve while the jet is launched, and what is the maximum spin parameter in this case? We try to answer these questions using the model proposed in this paper. 

A super-massive BH ($M \sim 10^9 M_{\odot}$) can be fed and spun up by accreting matter with a consistent sense of the angular momentum [the first calculations for a Kerr BH were performed by \citet{bardeen70}] or by merging with another BH \citep[e.g.,][]{berti,gergely-bp}. The general relativistic effects on the structure of the inner regions of an accretion disc surrounding a Kerr BH were first studied by \citet{nt} and \citet{pt} using Bardeen et al.'s (1972) orthonormal frames of the locally non-rotating observers. These studies resulted in a geometrically thin, optically thick accretion disc model [see also \citet{ss} for a quasi-Newtonian approach to the description of the disc accretion on to a Schwarzschild BH]. This is known as the standard, thin-accretion disc model. The model assumes that the disc is quasi-Keplerian (i.e., the radial pressure is negligible and the radial velocity of the flow is much smaller than its azimuthal component), and extends to the innermost stable (circular) orbit [see \citet{bardeenPT} for details on the orbits around a Kerr BH]. The disc is driven by internal viscous torque, which transports the angular momentum of the disc outwards, allowing the disc matter to be accreted on to the BH. It is assumed that the torque vanishes at the innermost stable orbit, so that the disc matter plunges into the BH carrying the specific energy and angular momentum that it has had at the innermost stable orbit. In the general relativistic regime, accretion on to the BH implies the conversion of the rest-mass energy of the infalling matter in the BH potential wall (i.e., the gravitational energy of the accretion disc) into kinetic and thermal energy of the accreted mass flow. If the thermal energy is efficiently radiated away, the orbiting gas becomes much cooler than the local virial temperature, and the disc remains geometrically thin. In the inner regions of the disc, the radiation pressure dominates the gas pressure. The opacity is dominated by electron scattering; i.e., the photons random-walk before leaving the disc as they scatter off of electrons. Since the half-thickness of the disc at a given radius, $r$, is much smaller than the radius itself, $h(r) \ll r$, the disc structure can be described through one-dimensional (1-D) hydrodynamic equations which are integrated in the vertical direction. \citet{db} developed a model of thin accretion disc driven jets, which can explain the shape of UV spectra from an AGN when the disc is sub-Eddington. Models of jet/wind formation from an accretion disc typically invoke specific magnetic field structures, as the synchrotron radio emission observed in (extra)galactic jets is possible only if magnetic fields are present. The jet can be launched and collimated, for instance, by centrifugal and magnetic forces \cite[e.g.,][]{bp82}. A possible condition for centrifugal launching of jets from a thin accretion disc is that the coronal particles, which are found just above the disc, should go into unstable orbits around the BH \citep{lyutikov09}. The jets/winds can also be launched from either (i) a geometrically thick disc with, e.g., an advection-dominated accretion flow \cite[ADAF, e.g.,][]{narayan94,armitage99}, a convection-dominated flow \citep{meier01}, or an advection-dominated inflow and outflow \citep{bb}, or (ii) a layer located between the accretion disc and the BH corona which consists of a highly diffusive, super-Keplerian rotating and thermally dominated by virial-hot and magnetised ion-plasma \citep{hujeirat02}. [See also the work by \citet{kuncic04} for the first fully analytic description of a turbulent magnetohydrodynamics accretion disc coupled to a corona.] \citet{fb95,fb99} proposed a jet-disc symbiosis model for powering jets; starting from the assumption that radio jets and accretion discs are symbiotic features present in radio quasars, these objects consist of a maximal jet power with a total equipartition (i.e., the magnetic energy flow along the jet is comparable to the kinetic jet power), and the total jet power is a particular fraction of the disc power. This fraction can be found by fitting the jet parameters to the observational data.

The energy and angular momentum of a BH can be electromagnetically extracted in the presence of a strong magnetic field threading the BH and supported by external currents flowing in the accretion disc, as shown by \citet{bz}. In this case, the energy flux of the jets is provided by conversion of the BH rotational energy into Poynting flux, which is then dissipated at large distances from the BH by current instabilities \citep{lyutikov02}. The Blandford--Znajek mechanism has been widely applied to jet formation in AGN, as well as to microquasar jets and gamma ray bursts, in an attempt to match a number of observational data. \citet{mt} explained this mechanism in terms of the BH membrane paradigm, in which case an imaginary stretched horizon (a conducting surface located just outside of the BH event horizon) mimics the BH electrodynamics as seen by outside observers. From the viewpoint of \citet{membrane}, the membrane paradigm implies not only the existence of a stretched horizon, but also a 3+1 split of space-time into (absolute) space and (universal) time. Therefore, the stretched horizon is regarded as a 2-D space-like surface that resides in 3-D space and evolves in response to driving forces from the external universe \citep{price86}. As a result, outside observers can make measurements at the stretched horizon and describe the physical properties of the horizon using pre-relativistic equations, such as Ohm's law. 

Different theoretical models of jet formation have been tested already by using numerical simulations. For instance, general relativistic magnetohydrodynamics (GRMHD) simulation results are consistent with models of gas pressure and magnetically driven jets \citep[e.g.,][]{shinjiet99,yosuke04,ken05,hawley06,mizuno07}, as well as with the Blandford--Znajek mechanism \citep[e.g.,][]{komissarov01,shinji03,mckinney04}. On the other hand, \citet{punsly09} performed fully relativistic 3-D MHD simulations of jets driven through the interaction of the magnetic field with the accreting gas in the BH ergosphere. The BH ergosphere is a part of the stationary asymptotically flat space-time (as the Kerr space-time) in which the Killing\footnote{Since the Kerr space-time is stationary (i.e., time-independent) and axially symmetric, there are two Killing vectors associated with these two symmetries \citep[e.g.,][]{kerr07}.} vector that corresponds asymptotically to time translation becomes space-like \citep{friedman78}. Therefore, negative energy states of matter that can extract energy from the BH are allowed there \citep{penrose69,bardeenPT}. The ergosphere lies outside of the BH event horizon, and its boundary intersects the event horizon only at the poles. At the stationary limit surface, an observer must move at the speed of light opposite the rotation of the BH just in order to stay still. Inside the ergosphere, the space-time itself is dragged in the direction of the BH rotation; i.e., nothing can stay there at rest with respect to distant observers, but it must orbit the BH in the same direction in which the BH rotates. This process is called the dragging of inertial frames \citep[e.g.,][]{mtw73}.

The BH-disc magnetic connection, first mentioned by Zel'dovich \& Schwartzman and quoted in \citet{t74}, can occur and change the energy-angular-momentum balance of the accreting gas in the disc \citep[e.g.,][]{mt,membrane,bland99,vanPutten99}. \citet{li00a,li00b,li02} derived the equations for the energy and angular momentum transferred from a Kerr BH to a geometrically-thin accretion disc (which consists of a highly-conducting ionised gas) by magnetic connection, and we shall use these equations. [See also the work by \citet{wang02}.] As the BH rotates relative to the disc, an electromotive force is generated. This drives a poloidal electric current flowing through the BH and the disc and produces an additional power on the disc. From the conservation laws of energy and angular momentum for a thin Keplerian accretion disc torqued by a BH, \citet{li02} calculated the radiation flux, the internal viscous torque and the total power of the disc, and found that the disc can radiate even without accretion. \citet{li02L} also looked for observational signatures of the BH-disc magnetic connection as more energy is radiated away from the disc and showed that the magnetic connection can produce a very steep emissivity compared to the standard, thin-accretion disc model. \citet{uzdensky04,uzdensky05} obtained the numerical solution of the Grad-Shafranov equation for a BH-disc magnetic-connection configuration in the case of both Schwarzschild and Kerr BHs. The Grad-Shafranov equation is a non-linear, partial differential equation that describes the magnetic flux distribution of plasma in an axisymmetric system. Uzdensky found that this BH-disc magnetic connection can only be maintained very close to the BH (see in the next section). In recent years, a number of models that also include the BH-disc magnetic connection have been developed. A BH magnetic field configuration with both open and closed magnetic field lines was considered by \citet{lei05}, who described the field configuration by the half-opening angle of the magnetic flux tube on the horizon, which is determined by the mapping relation between the angular coordinate on the BH horizon and the radial coordinated on the accretion disc. \citet{wang} proposed a toy model for the magnetic connection, in which case a poloidal magnetic field is generated by a single electric current flowing in the equatorial plane around a Kerr BH. \citet{ma07} derived the energy and angular momentum fluxes for a Kerr BH surrounded by an advection-dominated accretion disc. To solve the equations of the accretion flow, they used a pseudo-Newtonian potential. \citet{gan09} solved the dynamic equations for a disc-corona system and simulated its X-ray spectra by using the Monte Carlo method. \citet{zhao09} studied the magnetic field configuration generated by a toroidal distributed continuously in a thin accretion disc, as well as the role of magnetic reconnection in the disc to produce quasi-periodic oscillations in BH binaries. In the context of GRMHD, \citet{shinji06} presented a 2-D GRMHD result of jet formation driven by a magnetic field produced by a current loop near a rapidly-rotating BH, in which case the magnetic flux tubes connect the region between the BH ergosphere and a co-rotating accretion disc. Furthermore, relativistic Poynting jets driven from the inner region of an accretion disc that is initially threaded by a dipole-like magnetic field were studied by \citet{lovelace03}. Their model is derived from the special relativistic equation for a force-free electromagnetic field. [See also \citet{lynden-bell96,lynden-bell03}.] 

In this paper, we propose a model for launching relativistic jets from a (geometrically-thin) disc inside the ergosphere as an effect of the rotation of the space-time. We consider here the BH-disc magnetic connection, whose main role is to provide the source of energy for the jets when the mass accretion rate is very low. We use the general relativistic form of the conservation laws for the matter in a thin accretion disc to describe the disc structure when both the BH-disc magnetic connection and the jet formation are considered. The model is based on the calculations of \citet{nt}, \citet{pt} and \citet{li02}, being mainly influenced by the work of \citet{znajek78} and \citet{mt}. [Some incipient ideas which are at the base of this model were exposed in \citet{eu04,eu05}.] This is the first work that studies the process of jet launching from a geometrically-thin accretion disc inside the BH ergosphere when the energy and angular momentum are transferred from the BH to this region of the accretion disc via closed magnetic field lines, within the framework of general relativity. An important result of the model, with impact on observation of the AGN jets, is that the power of the jets does not depend linearly on the mass accretion rate all the way down to very low accretion rates for BHs of a given mass. This result is different from that of \citet{allen06}, who found a linear dependence between the power of the jet and the mass accretion rate by considering a spherical Bondi-type accretion on to BHs (in which case the accreting matter has zero or very low angular momentum). In their calculations, the power of the jet is estimated from the energy and time scale required to inflate the cavity observed in the surrounding X-ray emitting gas. The model proposed here combines two regimes associated with the driving of the jets, an accretion power regime and a (BH) spin-down power regime, where the switch from the former to the latter regime corresponds to a mass accretion rate of $\dot{m}\sim 10^{-1.8}$. In the accretion power regime, the power of the jets is linearly dependent on the mass accretion rate, whereas in the spin-down power regime the power of the jets depends very weakly on the mass accretion rate. In the accretion power regime, the energy and angular momentum are extracted and transported away from the disc inside the BH ergosphere by both the kinetic flux of particles and Poynting flux in the form of jets. Instead, in the spin-down power regime the energy and angular momentum are extracted and carried away from the disc inside the ergosphere predominantly in the form of Poynting flux with just little amount of kinetic flux of particles. The work presented in this paper is different from that of \citet{wang08}, in which the production of Poynting flux jets is associated with a combination of the Blandford--Znajek mechanism, the BH-disc connection and the Blandford--Payne mechanism. Furthermore, we argue that the accretion, which is initially at either close to the Eddington rate or at low rates, can be driven in a non-radiant, geometrically thin and quasi-Keplerian disc inside the BH ergosphere by the external jet torque. This is distinctly different from optically thin, advection-dominated accretion flow models \cite[e.g.,][]{narayan94}, in which the accretion at low rates is advection-dominated (i.e., the thermal energy generated via viscous dissipation is mostly retained by the accreted mass flow rather than being radiated, and the energy is advected in towards the BH) and quasi-radial (i.e., the flow is geometrically thick, having roughly a spherical structure). Our first goal in the present work is to obtain estimations for jet-related quantities, in particular the mass loading, power and Lorentz factor of the jet, when the BH rotational energy is transferred to the disc inside the ergosphere, and then to compare them with those derived from the Blandford-Znajek mechanism. In the limit of the spin-down power regime, the model proposed here can be regarded as a variant of the Blandford-Znajek mechanism, where the BH rotational energy is transferred to the disc inside the ergosphere rather than transported to remote astrophysical loads. Our second goal is to determine the upper limit of the spin parameter attained by a stationary Kerr BH when both jet formation and BH-disc magnetic connection are considered and to investigate how the value of the mass accretion rate can influence the departure of the BH spin parameter from its theoretical maximum limit of $a_* = 1$. 

In Section \ref{sec:assumptions}, we describe the assumptions of the model. In Section \ref{sec:qjet}, we derive the mass flow rate into the jets. Using the general-relativistic conservation laws for matter in the accretion disc (Section \ref{sec:conserv}), we derive the launching power of the jets (Section \ref{sec:jetspower}) and the angular momentum removed by the jets (Section \ref{sec:angmom}). In Section \ref{sec:efficiency}, we calculate the efficiency of launching the jets and show that when the BH accretes at low rates, the spin-down of the BH is an efficient mechanism to launch the jets via the accretion disc. In Section \ref{sec:evolution}, we study the spin evolution of the BH and discuss conditions of BH stationary states for given mass accretion rates. In Section \ref{sec:relevance}, we refer to the long lifetime of an AGN from the BH spin-down power as a particular relevance of the proposed model to the observational data. In Section \ref{sec:summary}, we present a summary of the key points, as well as our conclusions, and suggest further work to be done.

\section{Basic assumptions}
\label{sec:assumptions}

\begin{itemize}
\item We consider that matter outside of the BH has negligible gravitational effects compared to the BH gravity and that an accretion disc settles down in the equatorial plane of a Kerr BH. We assume that the accretion disc is geometrically thin and quasi-Keplerian, therefore the physical quantities integrated over the vertical direction can be used in order to describe the radial structure of the disc. The inner part of the accretion disc is located inside the BH ergosphere, extending from the stationary limit surface (or ergosphere) inward to the innermost stable orbit (see Fig. \ref{fig:jetBlack}). In the BH equatorial plane, the stationary limit surface is located at $r_{\rm sl} = 2\, r_{\rm g}$ and does not depend on the BH spin parameter, whereas the radius of the innermost stable orbit depends on $a_*$  [cf. eq. 2.21 of \citet{bardeen70}]; so that, $r_{\rm ms} (a_*) = 2\, r_{\rm g}$ for $a_* = 0.95$ and $\sim 1.2\, r_{\rm g}$ for $a_*\sim 1$. Here, $r_\mathrm{g}=GM/c^2$ is the gravitational radius [$r_{\mathrm{g}} = r_{\mathrm{g}}^{\dagger} (M/10^9 M_{\odot}) = 1.48 \times 10^{14} (M/10^9 M_{\odot})$ cm], $G$ is the Newtonian gravitational constant, $M$ is the BH mass and $c$ is the speed of light.

\begin{figure}
\epsfig{file= 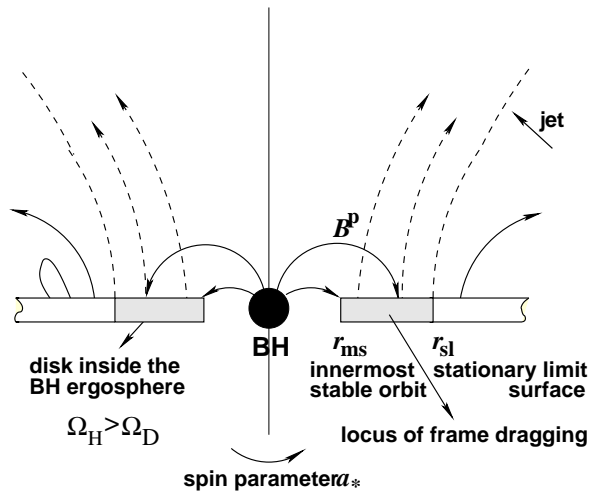,height=6.5cm}
\caption{Schematic representation of the inner part of the accretion disc-BH-jet system, where the BH is represented by the stretched horizon \citep{membrane}. Above the surface of the disc inside the BH ergosphere, the closed magnetic field lines (solid lines) do not cross the open magnetic field lines (dashed lines); they overlap only in the line-of-sight projection. For some explanation on the structure of the magnetic field in the disc inside the BH ergosphere, the reader is referred to the text below.}
\label{fig:jetBlack}
\end{figure} 

\item We adopt the viewpoint of the BH membrane paradigm of \citet{membrane} and represent the BH by the stretched horizon.

\item We consider the case of rapidly-spinning BHs with a spin parameter $a_* \geqslant 0.95$, based on the argument by \citet{bardeen221} that a strong preference for a particle to orbit in the equatorial plane requires the BH spin parameter to be close to its maximum value. 

\item We consider that closed magnetic field lines connect the BH to the accretion disc \cite[e.g.,][]{bland99,li00b}. The poloidal component of the magnetic field transfers angular momentum and energy (in the form of Poynting flux) from the BH to the disc, thereby increasing the amount of the gravitational energy which is released from the accretion disc. This energy is liberated very close to the BH, where most of the gravitational energy of the accretion disc is available (in our case, from the disc inside the BH ergosphere). Next, we assume that the disc energy (which, once again, represents the disc gravitational energy plus the BH rotational energy deposited into the disc by magnetic connection) is used to launch the jets. In this way, the disc remains cool and geometrically thin. By comparison, in a standard, thin accretion disc, the thermal energy is efficiently radiated away, so that the orbiting gas becomes much cooler than the local virial temperature, and the disc remains geometrically thin. 

\item As the numerical results obtained by \citet{uzdensky05} indicate, a closed magnetic field configuration can only be maintained in a region close to the BH. This region is limited to the radius of $\sim 12\,r_{\rm ms} = 72 \,r_{\rm g}$ for a static BH ($a_* = 0$) and decreases to $\sim 6\,r_{\rm ms} = 20.4\, r_{\rm g} $ as the BH spin parameter increases to $a_* = 0.7$. Therefore, it might be possible that, for a BH with $a_* \geqslant 0.95$, the closed magnetic field configuration to exist only in the region bounded by the BH ergosphere.

\item We constrain the jet formation to the disc inside the ergosphere, where the rotational effects of the space-time become much stronger. The slope of the specific energy of disc particles (Eq. \ref{eq:spenergy}) for $a_* \geqslant 0.95$ steepens in the BH ergosphere, indicating that most of the gravitational energy of the accretion disc is released from there. Moreover, inside the ergosphere, the space-time is dragged in the direction of the BH rotation and everything must co-rotate with the BH. (We shall ignore the possibility for the plasma inside the BH ergosphere to have negative mechanical energy at infinity.) Outside the ergosphere, it is plausible that patches of the disc gas may counter-rotate with the BH due to some instability or turbulence in the disc. If some magnetic field lines are frozen in such patches, they likely reconnect with the co-rotating magnetic field lines, and the closed magnetic field configuration can be destroyed. On the other hand, following the discussion above, it would appear that a closed magnetic field configuration will exist only in the region bounded by the BH ergosphere. And it is this magnetic field that transfers rotational energy from the BH to the accretion disc.

\item A more general magnetic field threading the BH would consist of a combination of closed magnetic field lines, as discussed above, and open magnetic field lines (extending from the BH to infinity), which can enable the extraction of BH rotational energy through the Blandford--Znajek mechanism. Now, one can wonder: (i) How much of the rotational energy of the BH is directly extracted by jets and how much is transferred to the disc, and (ii) under which specific conditions? These questions can be addressed in some future work. We limit the present work to the case of a BH threaded only by closed magnetic field lines and compare the values of some basic physical quantities (such as the power) of a jet driven solely from the disc inside the BH ergosphere with those of a jet powered by the Blandford-Znajek mechanism.

\item We assume that the jets remove angular momentum from the disc, enabling the accretion process. The external torques acting on the disc inside the BH ergosphere (i.e., magnetic BH-disc torque and jet torque) can dominate over the internal viscous torque  of the disc \citep{blandford01}, where the internal torque can be produced by magneto-rotational instability \cite[e.g.,][]{balbus91}, which is driven by the free energy available from the differential rotation of the gas flow.

\item To form the jet, at some point, the particles must cross the magnetic field lines. One can picture this as being due to drifts and instabilities inside the disc \citep{balbus94,balbus96,balbus98}. In addition to the closed magnetic field lines that connect the BH to the disc inside the ergosphere, the magnetic field can be represented similar to the magnetic field on the Sun surface. In a direct top-down view, it looks mottled, with arcs of magnetic fields connecting different regions and other magnetic flux tubes extending into free space and allowing matter to flow out [see also fig. 36 of \citet{membrane}]. [An accretion disc with a magnetic field configuration similar to that of the Sun was studied by \citet{yuan09} for driving episodic magnetohydrodynamic (MHD) jets.] However, the closed (BH-disc) and open (disc-infinity) magnetic field lines are supposed to be dominant in the model proposed here.

\item To examine the fate of the magnetic field in the space around the BH, we follow the ideas of \citet{membrane} on the `cleaning' of the magnetic field by the BH stretched horizon to maintain an ordered magnetic field [see fig. 36 of \citet{membrane}]. Since the magnetic field is frozen into the accretion disc, the field lines are transported in toward the BH by the accretion flow. Once the flow reaches the innermost stable orbit, it drops out of the disc and falls directly into the BH, and becomes causally disconnected from the field lines to which it was attached. However, the flux conservation law assures that the field lines, although disconnected from their source, will be pushed onto the BH by the Maxwell pressure of the adjacent field lines or thread through the near-vacuum region between the accretion disc and the BH or pushed back into the disc via Rayleigh-Taylor instabilities. If this cleaning works efficiently enough, an approximately stationary and axisymmetric magnetic field of the type described in the paragraph above can be formed [see also \citet{li00s}], otherwise the process discussed in this paper is not continuous and a different description than that presented here must be employed. Furthermore, our approach here is simple in the sense that with the assumption of an approximately stationary and axisymmetric magnetic field, the outflow near the launching region is thereby taken to be quasi-stationary ($\partial/\partial t =0$). Therefore, we are limited by not being able to specify time-dependent effects in the outflow. We concentrate here on the launching region of the jets, determining the mass flow rate into the jets and other conserved quantities of the outflow, and discuss the importance of the magnetic field in accelerating the jets on sub-parsec scales without solving the entire flow problem (see Section \ref{sec:jetspower}). 
 
\item To allow for jet formation, we assume that the surface density of the disc and the mean radial velocity combine themselves in such a way to keep their product constant over the disc inside the BH ergosphere (see Section \ref{sec:qjet}). This can be thought of as being an effect of the BH rotation. By comparison, in the model of \citet{bp82} the disc particles are driven upwards by the gradient of the pressure in the disc to fill the corona around the disc. The plasma should be sufficient to produce whatever charge and current densities are required for an MHD flow. The magnetic field lines that pass through the corona will be bent from near vertical to make a certain angle with the disc surface. If this angle is greater than $60^{\circ}$, the particles will fling outwards under centrifugal forces. Then, as magnetic stresses become important, the particles will be further accelerated by the gradient of the magnetic pressure. Far from the disc, the particles inertia will cause the magnetic field to become increasingly toroidal. As a result, the flow will be collimated by the magnetic hoop stresses to a cylindrical shape outside the (outer) light cylinder. In the model proposed here, the rotation of the space-time is responsible for ejecting particles in the direction perpendicular to the disc. The escape particles then slide along the open magnetic field lines (dashed lines in Fig. \ref{fig:jetBlack}), being accelerated by magnetic forces (see Section \ref{sec:jetspower}). 

\item Due to the motion of the disc plasma in which magnetic field lines are frozen, it is also possible that magnetic reconnection will take place at the interface of closed and open magnetic field lines. (Such a non-ideal MHD effect, which is common to impulsive and gradual flares, can convert magnetic energy into kinetic energy and thermal energy, and topologically change field lines; this can have significant consequences for the global evolution of a system. This effect is also connected with a violation of the magnetic flux conservation.) Such a magnetic reconnection that allows outflows to be driven has been observed in time-dependent resistive MHD simulations of jet/wind formation from neutron stars performed by \citet{romanova09}. This role of magnetic reconnection in driving outflows can also be extended to accreting BH systems. To estimate the rate of the magnetic energy that can be extracted through reconnection, we follow the work by \citet{dalpino05,dalpino10}. If we apply eq. 12 of \citet{dalpino05} to a BH of $10^9\,M_{\odot}$, we obtain a value of $\sim 10^{43}$ erg s$^{-1}$, which is about two orders of magnitude smaller than the minimum value of the power of the jet that we obtain in the model proposed here.
 
\item We use the Kerr metric \citep{kerr63} in cylindrical coordinates. In and near the BH equatorial plane, the metric is given by 
\begin{equation}
ds^{2}= -e^{2\nu}dt^2+e^{2\psi}\left( d\phi -\omega dt\right) ^2+e^{2\mu}dr^2+dz^2,
\label{eq:metric2}
\end{equation}  
where $r$, $\phi$ and $z$ are defined as the cylindrical coordinates in the asymptotic rest frame, and $t$ is the physical time of an observer removed to infinity \citep{pt}. The components of the metric tensor in (\ref {eq:metric2}) are specified by
\begin{equation}
e^{2\nu}=\frac{r^2\Delta}{A},\ e^{2\psi}=\frac{A}{r^2},\ e^{2\mu}=\frac{r^2}{\Delta},\ \omega =2r_\mathrm{g}aA^{-1},
\label{eq:coeff1}
\end{equation}
where the metric functions read
\begin{equation}
\Delta =r^2-2r_\mathrm{g}r+a^2, \ A=r^4+r^2a^2+2r_\mathrm{g}ra^2,
\label{eq:coeff2}
\end{equation} 
where $a=J/(Mc)$ is the angular momentum of the BH about the spinning axis ($J$) per unit mass and per speed of light. The BH spin parameter is defined as $a_* \equiv J/J{\max} \, (= a/r_{\mathrm{g}})$, where $J_{\max} = GM^2/c$ is the maximal angular momentum of the BH.
\end{itemize}

\section{Mass flow rate into the jets}
\label{sec:qjet}

\begin{figure}
\epsfig{file= 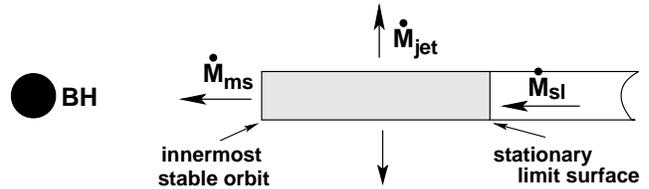,height=2.5cm}
\caption{Schematic representation of the mass flow through the disc inside the BH ergosphere.}
\label{fig:massflow}
\end{figure}

First, let us explain the terms used for the (rest) mass rates of the flow, which are measured by observers at infinity. The term mass accretion rate refers to the mass flow rate through the disc up to the ergosphere ($\dot{M}_{\rm D} = \dot{m}\dot{M}_{\rm Edd}$). In the ergosphere, $\dot{M}_{\rm D}$ is divided into the mass outflow rate into the jets ($\dot{M}_{\rm jets}$) and the mass inflow rate on to the BH ($\dot{M}_{\rm in}$). Now, we can write the mass outflow rate as
\begin{equation}
\dot{M}_{\mathrm{jets}}=\dot{M}_{\rm in}(\mathrm{r}_{\mathrm{sl}})-\dot{M}_{\rm in}(\mathrm{r}_{\mathrm{ms}}),
\label{eq:massjet}
\end{equation} 
where $\dot{M}_{\rm in}(\mathrm{r}_{\mathrm{sl}})$ and $\dot{M}_{\rm in}(\mathrm{r}_{\mathrm{ms}})$ denote the mass inflow rate at the stationary limit surface and at the innermost stable orbit, respectively. The mass outflow rate can also be expressed as
\begin{equation}
\dot{M}_{\mathrm{jets}} = q_{\mathrm{jets}} \dot{M}_{\mathrm{in}},
\label{eq:massjet3}
\end{equation} 
where the parameter $q_{\mathrm{jets}}$ indicates the fraction of the mass of the disc inside the BH ergosphere that goes into the jets. Figure \ref{fig:massflow} shows a schematic representation of the mass flow in the disc inside the ergosphere.

The amount of mass that flows inward across a cylinder of radius $r$ during a coordinate time interval $\Delta t$, when averaged by the method in \citet{nt}, is 
\begin{equation}
\dot{M} = -2\pi \sqrt{|g|}\, \Sigma\, \bar{v}^{\hat{r}}\,\mathcal{D}^{1/2},
\label{eq:massacc}
\end{equation} 
where $\sqrt{|g|}=e^{\nu+\psi+\nu}=r$ is the square root of the metric determinant (Eqs.~\ref{eq:metric2} and \ref{eq:coeff1}), $\Sigma = 2h<\rho_0>$ is the surface density of the disc (with $h$ being the half-thickness of the disc and $<\rho_0>$ the density of rest mass), $\bar{v}^{\hat{r}}$ is the mass-averaged radial velocity and $\mathcal{D}=(1-2/r_*+a_*^2/r_*^2)$ is one of the functions used to calculate general relativistic corrections to the Newtonian accretion disc structure. Here, $r_*=r/r_{\mathrm{g}}$ is the dimensionless radius.

Next, we estimate the mass inflow rate at one specific radius of the disc inside the BH ergosphere by using (\ref{eq:massacc}), so that the mass outflow rate becomes 
\begin{equation}
\dot{M}_{\mathrm{jets}}= \left[ -2\pi r\, \Sigma\, \bar{v}^{\hat{r}}\,\mathcal{D}^{1/2}\right] _{\mathrm{r}_{\mathrm{sl}}}-\left[ -2\pi r\, \Sigma\, \bar{v}^{\hat{r}}\,\mathcal{D}^{1/2}\right]_{\mathrm{r}_{\mathrm{ms}}}.
\end{equation} 

Now, the mass inflow rate at the stationary limit surface is also given by $\dot{m}\dot{M}_{\mathrm{Edd}}$, and then
\begin{equation}
\dot{M}_{\mathrm{jets}}= \dot{m}\dot{M}_{\mathrm{Edd}}\left[1-\frac{\dot{M}(\mathrm{r}_{\mathrm{ms}})}{\dot{M}(\mathrm{r}_{\mathrm{sl}})} \right].
\label{eq:massjet2}
\end{equation} 

By analogy with a standard, thin accretion disc [eqs. 5.9.5 and 5.9.10 of \citet{nt}], the product of the surface density and the mass-averaged radial velocity can depend on the disc radius as $\Sigma \bar{v}^{\hat{r}} \propto r^p$. One can recover a standard, thin accretion disc by setting $p=-1$, in which case the mass accretion rate [Eq. \ref{eq:massacc}, also eq. 5.6.2 of \citet{nt}] is independent of the radius. Now, instead of trying to obtain an exact condition for jet launching, let us simply take $p = 0$, as this gets rid of the requirement to know $\Sigma$ and $\bar{v}^{\hat{r}}$ precisely. Let us examine the consequences of this choice. Clearly, $p = 0$ corresponds to a product $\Sigma \bar{v}^{\hat{r}}$ constant for any radius of the disc inside the ergosphere, and consequently the mass inflow rate on to the BH decreases with radius as $\dot{M}_{\mathrm{in}}(r) \sim r$, and the difference flows into the jets. Then, $\dot{M}_{\mathrm{in}}(r)$ at a given radius is specified by the general relativistic factor $\mathcal{D}^{1/2}$, which depends on the BH spin parameter. For an ADAF disc that allows outflows, $\dot{M}_{\mathrm{in}}(r) \sim r^{q}$, where $0\leqslant q < 1$ [eq. 13 of \citet{blandford99}]. Therefore, our choice of $q=1$ (or $p = 0$ ) represents a limiting case in the model by \citet{bb}.

\begin{figure}
\centering
\epsfig{file= 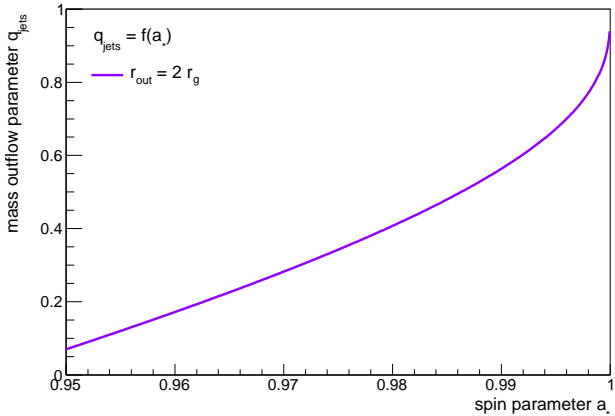,height=5.5cm}
\caption{Mass outflow parameter ($q_{\mathrm{jets}}$) as a function of the BH spin parameter ($a_*$). For $a_* \sim 1$, almost whole material of the disc inside the BH ergosphere flows into the jets ($q_{\mathrm{jets}}\simeq 0.98$); that is, the BH almost stops being fed by accreting matter.}
\label{fig:qjets}
\end{figure}

Using equations (\ref{eq:massjet3}) and (\ref{eq:massjet2}), as well as the expression of $\mathcal{D}$, we obtain the fraction of the mass inflow that goes into the jets (or the mass outflow parameter) as
\begin{equation}
q_{\mathrm{jets}}(a_*) = 1-\frac{r_{\mathrm{ms}_*}}{r_{\mathrm{sl}_*}}\,\left( \frac{1-2/r_{\mathrm{ms}_*}+a_*^2/r^2_{\mathrm{ms}_*}}{1-2/r_{\mathrm{sl}_*}+a_*^2/r^2_{\mathrm{sl}_*}}\right) ^{1/2},
\label{eq:qjets}
\end{equation} 
where $r_{\mathrm{ms}_*}=r_{\mathrm{ms}}/r_{\mathrm{g}}$ and $r_{\mathrm{sl}_*}=r_{\mathrm{sl}}/r_{\mathrm{g}}$.

Figure \ref{fig:qjets} shows the mass outflow parameter as a function of the BH spin parameter. For $a_* = 0.95$, the mass outflow into the jets is only about 8 per cent of the available mass inflow through the disc inside the ergosphere. Instead, for a spin parameter near the maximal one ($a_* \sim 1$), the mass outflow increases to about 98 per cent of the available mass inflow. This means that in the case of near maximal spin, the BH almost stops being fed by accreting matter. 

Suppose the inner disc would have been extended beyond the stationary limit surface. In this case, the disc particles can form the jets ($q_{\mathrm{jets}} > 0$) if and only if the BH spin parameter were $a_* > 0.755$ (plot not shown). 

We mention that the results presented in this section are valid for our choice of $p = 0$. This, of course, need not be a necessary condition for jet launching, since we have examined the mass outflow parameter for just one value of $p$, that which makes $\Sigma \bar{v}^{\hat{r}}$ constant for any radius of the disc inside the BH ergosphere and $q_{\mathrm{jets}}$ dependent only on the BH spin parameter, but it is certainly sufficient. The BH rotation causes an outflow of particles from the disc, where the energy (and angular momentum) carried by the escape particles is taken from the accretion disc. The escape particles then slide along the open magnetic field lines, being accelerated by magnetic forces (see Section \ref{sec:jetspower}).

\section{Angular momentum and energy conservation laws}
\label{sec:conserv}

To describe the structure of the disc inside the BH ergosphere, we use the angular momentum and energy conservation laws derived by \citet{pt} and include both the BH-disc magnetic connection and the jet formation. When deriving the conservation laws, \citet{pt} do not make any assumption about the type of stress-energy present (e.g., magnetic fields, viscous stresses, etc.). [The calculations performed by \citet{pt} are valid even if the disc is highly dynamical, but can also be applied to steady-state and quasi-steady-state discs, in which case the mass accretion rate is constant throughout the disc.] Here, we consider that the removal of the angular momentum of the disc inside the ergosphere can be produced by the external jet torque and that the external torques acting on the disc inside the ergosphere (i.e., BH-disc magnetic torque and jet torque) dominate over the internal viscous torque of the disc \citep{blandford01}. In this case, we can write the angular momentum conservation law\footnote{The equation describing the angular momentum conservation is derived in Appendix \ref{conslaws}.} as
\begin{equation}
\frac{d}{dr}\left[ \left( 1-q_{\mathrm{jets}}\right) {\dot{M}}_{\mathrm{D}}cL^{\dagger}\right]+4\pi rH =4\pi rJL^{\dagger}, 
\label{eq:angmom}
\end{equation} 
where on the left-hand side, the first term describes the angular momentum carried by the accreting mass of the disc inside the BH ergosphere, and the second term describes the angular momentum transferred from the BH to the disc inside the ergosphere. The term on the right describes the angular momentum carried away by the jets. $L^{\dagger}$ is the specific angular momentum of a gas particle orbiting in the accretion disc, $J$ is the total flux of energy (of particle and electromagnetic origin) carried away by jets and $H$ is the flux of angular momentum transferred from the BH to the disc inside the ergosphere. $H$ is defined through the magnetic torque produced by the BH on both surfaces of the accretion disc $T_{\mathrm{HD}}$ \citep{li02} 
\begin{equation}
T_{\mathrm{HD}}= 2\int_{r_1}^{r_2}2\pi rHdr,
\label{torque}
\end{equation}
where the limits of integration are two radii of the accretion disc with $r_1 < r_2$. 

Similar to the angular momentum conservation law, we can write the energy conservation law as 
\begin{equation}
\frac{d}{dr}\left[ \left( 1-q_{\mathrm{jets}}\right) {\dot{M}}_{\mathrm{D}}c^2E^{\dagger}\right]+4\pi r H \Omega_{\mathrm{D}}  =4\pi rJE^{\dagger}, 
\label{eq:energy}
\end{equation} 
where on the left hand-side, the first term describes the rate of the energy flow through the disc inside the BH ergosphere, and the second term is the rate at which the magnetic torque per unit area of the disc does work, $T_{\mathrm{HD}}\Omega_{\mathrm{D}}$ (here $\Omega_{\mathrm{D}}$ is the Keplerian angular velocity of the gas particles in the disc). The third term describes the energy flow along the jets. $E^{\dagger}$ is the specific energy of a gas particle having mass $\mu$ and orbiting in the same direction as the BH rotation \citep{bardeenPT}:
\begin{equation}
E^{\dagger} \equiv \frac{E}{\mu}=\frac{r^{3/2}-2r_{\mathrm{g}}r^{1/2}+r_{\mathrm{g}}^{1/2}a}{r^{3/4}\left(r^{3/2}-3r_{\mathrm{g}}r^{1/2}+2r_{\mathrm{g}}^{1/2}a\right)^{1/2} }.
\label{eq:spenergy}
\end{equation} 

The flux of angular momentum transferred from the BH to the disc inside the ergosphere by magnetic connection has the following expression \citep{li02}:
\begin{equation}
H=\frac{1}{8\pi^{3}r}\left( \frac{d\Psi_{\mathrm{D}}}{c\,dr}\right) ^{2}\frac{\Omega_{\mathrm{H}}-\Omega_{\mathrm{D}}}{\left(-dR_{\mathrm{H}}/dr\right) },
\label{eq:li}
\end{equation}
where $\Psi_{\mathrm{D}}$ is the flux of the poloidal magnetic field lines which thread the surface of the disc inside the BH ergosphere and $\Omega_{\mathrm{H}}$ is the BH angular velocity. The derivation of Eq. (\ref{eq:li}) is based on the supposition that the accretion disc consists of a highly conducting ionised gas. This implies that (i) the accretion disc resistance is neglected in comparison with the BH surface resistance and (ii) the magnetic field lines are frozen in the accretion disc, being transported by the disc gas and rotating with $\Omega_{\mathrm{D}}$. On the other hand, the angular velocity of the magnetic field lines threading the horizon is $\Omega_{\mathrm{H}}$, due to the effect of the frame-dragging at the BH horizon. For $a_* > 0.35$ and $r\geq r_{\mathrm{ms}}$, $\Omega_{\mathrm{H}} > \Omega_{\mathrm{D}}$, so that the BH transfers energy (and angular momentum) to the disc. For $a_* < 0.35$, $\Omega_{\mathrm{H}} < \Omega_{\mathrm{D}}$, and this time the accretion disc transfers energy (and angular momentum) to the BH. For $a_* = 0.35$, $\Omega_{\mathrm{H}} = \Omega_{\mathrm{D}}$; this condition implies that there is no energy (nor angular momentum) transfer between the BH and the accretion disc by magnetic connection.

\section{\label{sec:jetspower}Launching power of the jets}
\label{sec:jetspower}

We are now in the position to calculate the launching power of the jets with the help of the conservation laws previously derived. First, we define the launching power of both jets as
\begin{equation}
P_{\mathrm{jets}}=2 \int_{r_{\mathrm{ms}}}^{r_{\mathrm{sl}}}2\pi JE^{\dagger}rdr.
\end{equation}
Integrating the equation of the energy conservation law (Eq. \ref{eq:energy}) over the disc inside the BH ergosphere, we find the launching power of the jets,
\begin{equation}
P_{\mathrm{jets}}=\left( 1-q_{\mathrm{jets}}\right) {\dot{M}}_{\mathrm{D}}c^2\left( E^{\dagger}_{\mathrm{sl}} -E^{\dagger}_{\mathrm{ms}} \right)+4\pi \int_{r_{\mathrm{ms}}}^{r_{\mathrm{sl}}}rH\Omega_{\mathrm{D}}dr.
\end{equation} 
The first term describes the rest energy of the accreting matter on to the BH and the second term describes the energy transfer from the rotating BH to the disc inside the ergosphere. $E^{\dagger}_{\mathrm{sl}}$ and $E^{\dagger}_{\mathrm{ms}}$ are the specific energy of the gas particle (Eq. \ref{eq:spenergy}) evaluated at the stationary limit surface and at the innermost stable orbit, respectively.

Using Eq. (\ref{eq:li}), we obtain the launching power of the jets as
\begin{equation}
\begin{split}
P_{\mathrm{jets}} &= \left( 1-q_{\mathrm{jets}}\right){\dot{M}}_{\mathrm{D}}c^2\left( E^{\dagger}_{\mathrm{sl}}-E^{\dagger}_{\mathrm{ms}} \right)  \\
&+\frac{1}{2\pi^{2}}\int_{r_{\mathrm{ms}}}^{r_{\mathrm{sl}}}\left(\frac{d\Psi_{\mathrm{D}}}{c\,dr}\right)^{2}\,\frac{\Omega_{\mathrm{H}}-\Omega_{\mathrm{D}}}{\left( -dR_{\mathrm{H}}/dr\right)}\Omega_{\mathrm{D}}dr,
\end{split}
\label{eq:jetspow}
\end{equation} 
where the angular velocities of the BH and the accretion disc, respectively, are 
\begin{equation}
\Omega_{\mathrm{H}} \equiv \frac{c}{2r_{\mathrm{g}}}\,\frac{a_*}{1+\left( 1-a_*^2\right) ^{1/2} } = \frac{c}{r_{\mathrm{g}}}\,\Omega_{\mathrm{H*}},
\label{eq:omegaH} 
\end{equation}
\begin{equation}
\Omega_{\mathrm{D}} \equiv \frac{c}{r_{\mathrm{g}}}\,\frac{1}{r_*^{3/2}+a_*} = \frac{c}{r_{\mathrm{g}}}\,\Omega_{\mathrm{D*}} .
\label{eq:omegaD}
\end{equation}

To calculate the launching power of the jets, we need to evaluate both $\Psi_{\mathrm{D}}$ and $( -dR_{\mathrm{H}}/dr)$. First, we write the magnetic flux that threads the accretion disc surface,
\begin{equation}
\Psi_{\mathrm{D}}=\int B_{\mathrm{D}}(dS)_{z=0},
\label{eq:18}
\end{equation} 
where $B_{\mathrm{D}}$ is the poloidal component of the magnetic field that threads the disc. The surface area between two equatorial surfaces in a Kerr space-time can be calculated from
\begin{equation}
(dS)_{z=0}=\sqrt{\det g_{(r\phi)}}\,dr\,d\phi,
\label{eq:arie}
\end{equation}
where the determinant of the surface metric is
\begin{equation}
\det g_{(r\phi )}= \left| \begin{array}{cc} g_{rr} & g_{r\phi} \\
g_{\phi r} & g_{\phi \phi} \end{array} \right|=
\left| \begin{array}{cc} e^{2\mu} & 0 \\
0 & e^{2\psi} \end{array} \right|=
\frac{A}{\Delta}.
\end{equation}
This result follows from Eqs. (\ref{eq:metric2}), (\ref{eq:coeff1}), and (\ref{eq:coeff2}). With these, the surface area in Eq. (\ref{eq:arie}) reads 
\begin{equation}
(dS)_{z=0} = \left( \frac{A}{\Delta}\right) ^{1/2}2\pi\,dr.
\end{equation}

The poloidal component of the magnetic field that threads the BH horizon $B_{\mathrm{H}}$ and the poloidal component of the magnetic field at the inner edge of the accretion disc $B_{\mathrm{D}}(r_{\mathrm{ms}})$ can be of the same order \cite[e.g.,][]{livio} and related by 
\begin{equation}
B_{\mathrm{H}} = \zeta B_{\mathrm{D}}(r_{\mathrm{ms}})\,,\;\textrm{where}\;\zeta\geq 1.
\end{equation} 
On the other hand, the poloidal component of the magnetic field that threads the accretion disc surface scales as $B_{\mathrm{D}}\propto r^{-n}$, where $0 < n < 3$ \citep{bland76}. Consequently,
\begin{equation}
B_{\mathrm{D}} = B_{\mathrm{D}}(r_{\mathrm{ms}})\left( \frac{r}{r_{\mathrm{ms}}}\right) ^{-n}=\frac{B_{\mathrm{H}}}{\zeta}\,\left( \frac{r}{r_{\mathrm{ms}}}\right) ^{-n}.
\label{eq:polo}
\end{equation} 

Since the BH horizon behaves, in some aspects, like a rotating conducting surface \cite[e.g.,][]{carter73,damour,znajek78,membrane}, it can be thought of as being a `battery' driving currents around a circuit. The energy for this comes from the BH rotation \citep{znajek78}. The internal resistance of the battery in the horizon, i.e., the resistance between two magnetic surfaces that thread the horizon, is
\begin{equation}
dR_{\mathrm{H}}=R_{\mathrm{H}}\frac{dl}{2\pi r_{\mathrm{H}}},
\label{eq:rez}
\end{equation} 
where $R_{\mathrm{H}}=4\pi/c =377$ ohm, $dl$ is the horizon distance between two magnetic surfaces (see Fig. \ref{fig:jetBlack}), $2\pi r_{\mathrm{H}}$ is the cylindrical circumference at $r=r_{\mathrm{H}}$ and $r_{\mathrm{H}}=r_{\mathrm{g}}[1+(1-a_*^2)^{1/2}] = r_{\mathrm{g}} r_{\mathrm{H*}}$ is the radius of the BH horizon \citep{membrane}.

The voltage difference generated by the BH has a maximum magnitude of $V = \Omega_{\mathrm{H}} \Psi_{\mathrm{H}}$, where $\Psi_{\mathrm{H}} = B_{\mathrm{H}}A_{\mathrm{H}}$ is the magnetic flux threading the BH and $A_{\mathrm{H}} = 8\pi r_{\mathrm{g}}r_{\mathrm{H}}$ is the surface area of the BH. Assuming that the magnetic field is carried into the BH by the accreted disc gas, we set the BH potential drop to the energy of the gas particles carried into the BH, the latter being the particle specific energy at the innermost stable orbit. Suppose that during a first epoch, the BH accretes at a rate approximately equal to the Eddington rate.\footnote{The Eddington accretion rate is defined from the Eddington luminosity as ${\dot{M}}_{\mathrm{Edd}} = L_{\mathrm{Edd}}/(\varepsilon c^2) = 4\pi GM/(\varepsilon \kappa_{\mathrm{T}} c)$, where $\varepsilon$ is the efficiency of converting the gravitational energy of the accretion disc into radiation and $\kappa_{\mathrm{T}}$ denotes the Thomson opacity. $\varepsilon$ depends on the BH spin parameter as $\varepsilon=1-E^{\dagger}_{\mathrm{ms}}$ \citep{t74}, so that $\varepsilon= 0.06$ for a Schwarzschild BH and $\varepsilon= 0.42$ for an extremely spinning Kerr BH. We scale the BH mass to $10^9 M_{\odot}$, so that ${\dot{M}}_{\mathrm{Edd}} = {\dot{M}}_{\mathrm{Edd}}^{\dagger}\varepsilon^{-1}(M/10^9 M_{\odot})$, where ${\dot{M}}_{\mathrm{Edd}}^{\dagger} = 1.38\times 10^{26}$ g s$^{-1}$.} This supposition provides $V^2 = {\dot{M}}_{\mathrm{acc}}E^{\dagger}_{\mathrm{ms}}c^2$. Therefore, the maximum value of the magnetic field that threads the BH horizon is 
\begin{equation}
\left( B_{\mathrm{H}}^{\mathrm{max}}\right)^2=\frac{{\dot{M}}_{\mathrm{Edd}}c E^{\dagger}_{\mathrm{ms,lim}}} {4\pi r_{\mathrm{g}}^2 (a_{\mathrm{*,lim}} )^2},
\label{eq:magn}
\end{equation} 
where $a_{\mathrm{*,lim}} = 0.9982 $ is the BH limiting spin in the case of a radiatively efficient accretion disc \citep{t74}, and the corresponding particle specific energy at the innermost stable orbit is $E^{\dagger}_{\mathrm{ms,lim}}= 0.6759$. Although this limit of the BH spin may be even closer to its maximum value $a_* \sim 1$, it produces a negligible variation in the maximum value of the BH magnetic field. Using the expression of the gravitational radius, the maximum value of the magnetic field that threads the BH horizon (Eq. \ref{eq:magn}) becomes
\begin{equation}
\left( B_{\mathrm{H}}^{\mathrm{max}}\right)^2=\frac{{\dot{M}}_{\mathrm{Edd}}^{\dagger} c E^{\dagger}_{\mathrm{ms,lim}} }{4\pi \,\varepsilon_{\mathrm{lim}} (r_{\mathrm{g}}^{\dagger})^2 (a _{*, {\mathrm{lim}}})^2}\left( \frac{M}{10^9 M_{\odot}} \right)^{-1},
\label{BHfield}
\end{equation}
or 
\begin{equation}
B_{\mathrm{H}}^{\mathrm{max}} =  0.56 \times 10^4 \left(\frac{M}{10^9M_{\odot}}\right) ^{-1/2}\ \mathrm{gauss}.
\label{Bbhmax}
\end{equation}  
This result is similar to the calculation performed by \citet{znajek78}. [See also \citet{lovelace}.] The only difference is that we set the BH potential drop to the specific energy of the particles at the innermost stable orbit, whereas \citet{znajek78} makes use of the fact that the Eddington luminosity sets an upper bound on the radiation pressure (as the disc is radiatively efficient), and thus $V^2 \sim L_{\mathrm{Edd}}$.

The continuum of the magnetic field within a narrow strip between two magnetic surfaces, which connect the BH to the disc inside the ergosphere, $d\Psi_{\mathrm{H}}=d\Psi_{\mathrm{D}}$  \cite[e.g.,][]{wang}, gives
\begin{equation}
B_{\mathrm{H}}2\pi r_{\mathrm{H}}\,dl = - B_{\mathrm{D}}\left( \frac{A}{\Delta}\right) ^{1/2}2\pi \,dr.
\label{eq:cont}
\end{equation} 
Making use of Eqs. (\ref{eq:polo}), (\ref{eq:rez}) and (\ref{eq:cont}), we obtain
\begin{equation}
(-dR_{\mathrm{H}}/dr)=\frac{2}{c\,r_{\mathrm{H}}^2}\,\frac{1}{\zeta}\left( \frac{r}{r_{\mathrm{ms}}}\right)^{-n}\,\left(\frac{A}{\Delta} \right) ^{1/2}.
\label{dRdr}
\end{equation} 

Next, we write the particle specific energy (Eq. \ref{eq:spenergy}) using the dimensionless radius and the spin parameter. Then, we substitute this equation (evaluated at $r_{\mathrm{sl}_*}$ and $r_{\mathrm{ms}_*}$, respectively) together with (\ref{eq:omegaH}) -- (\ref{eq:magn}), (\ref{eq:cont}) and (\ref{dRdr}) for Eq. (\ref{eq:jetspow}). So, the launching power of the jets becomes
\begin{equation}
\begin{split}
P_{\mathrm{jets}}&=\dot{m}{\dot{M}}_{\mathrm{Edd}}^{\dagger}c^2\varepsilon^{-1}(1-q_{\mathrm{jets}})\left( E^{\dagger}_{\mathrm{sl}_*}-E^{\dagger}_{\mathrm{ms}_*}\right)\left(\frac{M}{10^9M_{\odot}}\right) \\  
&+{\dot{M}}_{\mathrm{Edd}}^{\dagger}c^2\, C_* \left( \frac{B_{\mathrm{H}}}{B_{\mathrm{H}}^{\mathrm{max}}}\right)^2\left(\frac{M}{10^9M_{\odot}}\right)  \\ &\int_{r_{\mathrm{ms}_*}}^{r_{\mathrm{sl}_*}}r_*^{1-n}\mathrm{R}_*^{1/2}\left( \Omega_{\mathrm{H}_*}-\Omega_{\mathrm{D}_*}\right) \Omega_{\mathrm{D}_*}dr_*,
\end{split}
\label{eq:jets}
\end{equation} 
where
\begin{equation}
\mathrm{C}_*=\frac{r_{\mathrm{H}_*}^2r_{\mathrm{ms}_*}^n E^{\dagger}_{\mathrm{ms}_*,{\mathrm{lim}}} }{4\pi \zeta (a_{*,\mathrm{lim}})^2 \varepsilon_{\mathrm{lim}} }, \;\mathrm{R}_*=\frac{1+a_*^2r_*^{-2}+2a_*^2r_*^{-3}}{1-2r_*^{-1}+a_*^2r_*^{-2}},
\label{eq:constante}
\end{equation} 
\begin{equation}
E^{\dagger}(r_*, a_*)=\frac{1-2r_*^{-1}+a_*r_*^{-3/2}}{\left( 1-3r_*^{-1}+2a_*r_*^{-3/2}\right) ^{1/2}}\,.
\label{eq:Edagger}
\end{equation} 
For the following calculations, we consider the strength of the magnetic field in Eq. (\ref{eq:jets}) to be as high as its maximum value $B_{H} \sim B_{\mathrm{H}}^{\mathrm{max}}$. On the right-hand side, the first term represents the accretion power of the disc inside the BH ergosphere and the second term represents the BH spin-down power transferred to the disc by magnetic connection. So, Eq. (\ref{eq:jets}) can also read:
\begin{equation}
P_{\mathrm{jets}}=P_{\mathrm{jets}}^{\mathrm{acc}}+P_{\mathrm{jets}}^{\mathrm{rot}}.
\end{equation}

\begin{figure}\centering
\epsfig{file= 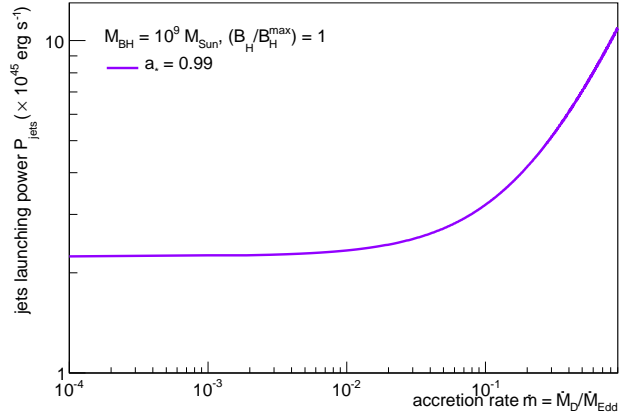,height=5.5cm}
\caption{Launching power of the jets as a function of the mass accretion rate $\dot{m}$ (Eq. \ref{eq:jets}) for a given BH spin parameter, $a_*=0.99$. The switch from an accretion power regime to a spin-down power regime corresponds to a mass accretion rate of $\dot{m} \simeq 10^{-1.8}$. }
\label{PjetsVsMdot}
\end{figure}

In Fig. \ref{PjetsVsMdot}, we plot the launching power of the jets as a function of the mass accretion rate. The plot shows that a transition from an accretion power regime to a spin-down power regime is produced for $\dot{m} \simeq 10^{-1.8}$. So, we have: (1) an accretion power regime in which case $\dot{m} > 10^{-1.8}$ and the dominating term in the launching power of the jets is $P_{\mathrm{jets}}^{\mathrm{acc}}$, and (2) a spin-down power regime in which case $\dot{m} < 10^{-1.8}$ and the dominating term in the launching power of the jets is $P_{\mathrm{jets}}^{\mathrm{rot}}$.

In Eq. (\ref{eq:jets}), the launching power of the jets depends on: (i) the mass accretion rate $\dot{m}$, (ii) the BH mass $M$, (iii) the BH spin parameter $a_*$, (iv) the power-law index n, and (v) the ratio of the magnetic field strengths $\zeta$. We chose the last two parameters as follows: the power-law index $n$ is taken to be `2' as for a frozen magnetic field \citep{alfven}, and $\zeta$ is set by taking its value corresponding to the maximum of the launching power of the jet, which is one. Therefore, for the following calculations, we consider $n = 2 \: \textrm{and} \,\zeta=1$.

\begin{figure}\centering
\epsfig{file=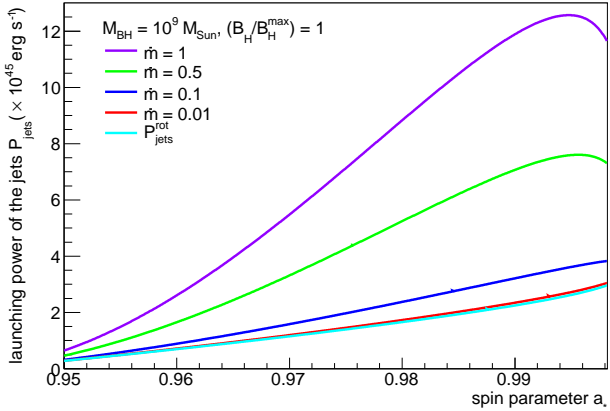,height=5.5cm}
\caption{Launching power of the jets as a function of $a_*$ (Eq. \ref{eq:jets}) for four values of the mass accretion rate. The mass accretion rates from top to bottom are 1, 0.5, 0.1 and 0.01. The bottom curve represents the power of the jets given by the BH spin-down $P_{\mathrm{jets}}^{\mathrm{rot}}$. Note that there is a slight difference between the two bottom curves (red and turquoise curves). In the case of very low mass accretion rates, $\dot{m} < 0.01$, $P_{\mathrm{jets}}$ is approximately equal to the BH spin-down power.}
\label{fig:Pjets}
\end{figure}

In Fig. \ref{fig:Pjets}, we plot the launching power of the jets as a function of the BH spin parameter, for a BH mass of $10^{9} M_{\odot}$, given four values of the mass accretion rate $(\dot{m}=1, 0.5, 0.1\,\textrm{and}\, 0.01)$, as well as the BH spin-down power contribution to the jets power (bottom curve). Since the area of the disc inside the ergosphere increases with an increase of $a_*$, there is a dominating trend of the jet power to increase as well, except for $a_*$ close to the maximal value and $\dot{m} > 0.1$ where the turn-over of the curve is produced due to the general relativistic factor that appears in the term $(1-q_{\mathrm{jets}})$ of the accretion power. In the case of the spin-down power regime, the jet power is $\sim 10^{45}$ erg s$^{-1}$, which is only $10^{-2}$ of the Eddington luminosity of a $10^{9}$ solar mass BH. This value of $P_{\mathrm{jets}}^{\mathrm{rot}}$ is comparable to the maximum rate of energy extraction by the Blandford--Znajek mechanism, which is $\sim 10^{45}$ erg s$^{-1}$ for a BH mass of $10^{9} M_{\odot}$ and $a_*$ close to the maximal value [Eq. 4.50 of \citet{membrane}]. For a lower mass of the BH, the jet power decreases, as the launching power of the jets is proportional to the BH mass.

On sub-parsec scales, the jets are likely to be dominated by electromagnetic processes (MHD or pure electrodynamic), where the energy is transported along the jets via Poynting flux, and are potentially unstable if significant thermal mass load is present \citep{meier03}. Next, we estimate the magnetization parameter of the jet plasma at the launching points, $\sigma$, which reflects the effect of a rotating magnetic field on accelerating the jet plasma by measuring the Poynting flux in terms of particle flux \citep[e.g.,][]{michel69,camenzind86,fendt01}. The initial magnetization parameter of the jets (denoted by an index `0') is given by
\begin{equation} 
 \sigma_0 = \frac{\Psi^2 \Omega_{\rm D}^{2}}{4\pi\dot{M}_{\rm jets}c^3},
\label{sigma}
\end{equation}
\begin{figure}
\epsfig{file= 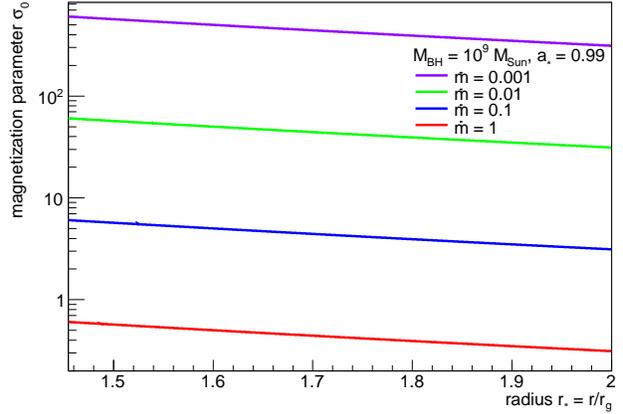,height=5.5cm}
\caption{Initial magnetization parameter of the jet plasma as a function of the disc radius for three values of the mass accretion rate when the BH spin parameter is $a_* = 0.99$. The mass accretion rates from top to bottom are 0.001, 0.01, 0.1 and 1. For $\sigma \geqslant 1$, the jets are Poynting flux-dominated outflows; i.e., the energy content of the jets is mainly in the magnetic part. For $\sigma < 1$, the jets are particle-dominated outflows. Note that $\sigma_0 \sim \dot{m}^{-1}$.}
\label{sigma99} 
\end{figure} 
where $\Psi = \int B\,dS$ and $ \Omega_{\rm D}$ is here taken as the angular velocity of the magnetic field lines frozen in the disc. For illustration, we evaluate $\sigma_0$ for three values of the mass accretion rate ($\dot{m} = 1, 0.1, 0.01\,\textrm{and}\,0.001$) in the case of a BH with the spin parameter of $a_* = 0.99$ (see Fig. \ref{sigma99}). The magnetization parameter increases with decreasing mass accretion rate ($\sigma_0 \sim \dot{m}^{-1}$), as well as with decreasing radius. When $\sigma_0 > 1$, the Poynting flux dominates in the jets and the energy can be transferred from the magnetic field to the particles. As a result, the jets can be accelerated on the expense of the stored energy in the magnetic field as the Poynting flux is converted into kinetic energy flux by magnetic forces \cite[e.g.,][]{fendt04}. The magnetic force can be split in two components: the magnetic pressure force ($\sim \nabla B_{\phi}^2$ ), which points in positive outward direction, and the magnetic tension force ($\sim B_{\phi}^2$ ), which points in negative inward direction. ($B_{\phi}$ denotes the strength of the toroidal component of the magnetic field.) In the case of a ballistic jet, which expands with a constant speed, the two forces cancel each other and the toroidal component of the magnetic field decreases as $B_{\phi} \sim z^{-1}$ due to the magnetic flux conservation. ($z$ denotes the distance along the jet.) To accelerate the flow, the magnetic pressure gradient must prevail over the magnetic tension force. This can be possible, in principle, due to the decrease in the strength of the toroidal magnetic field as the jet propagates away from the source, but one has to solve the full problem to verify whether the magnetic pressure gradient dominates over the magnetic tension force. The magnetic acceleration process is limited by the free energy available in the magnetic field, and saturation must occur at some point. When the kinetic energy flux becomes dominant in the jet, strong shocks can occur. These shocks can further accelerate the jet. However, if the magnetic field becomes highly twisted, the magnetic field itself will not be able to explain the acceleration of the jet, as the magnetic reconnection or other instability in the jet can lead to magnetic energy dissipation and shock formation. (Strong toroidal magnetic fields are subject to the kink instability, which excites large-scale helical motions that can distort or even disrupt the jet [e.g., \citet{mizuno09} and references therein]. However, the growth rate of the kink mode may be reduced, for instance, by increasing the magnetic pinch or by including a gradual shear, an external wind or relativistic bulk motion.) How far the magnetic acceleration of the jets can occur depends from one case to another, and the ambient medium into which jets propagate can play a significant role. On parsec scales, relativistic shocks are expected to be prominent \citep{lobanov07}. A first stationary, strong shock can be produced in the approximate range $(3-6)\times 10^3 \, r_{\rm g}$  \citep{markoff01,marscher}, whereas moving shocks can occur between 20 and 200 $r_{\mathrm{g}}$. For very large values of the magnetization parameter, the MHD approximation breaks down \citep{mizuno09}. For  $\sigma_0 = 1$ the Poynting flux and the particle kinetic energy flux are in equipartition, whereas for $\sigma_0 < 1$ the jets are kinetic energy flux-dominated outflows. For given mass accretion rate and BH spin parameter, the value of $\sigma_0$ is not much less than one, being only within one order or magnitude smaller than one. Therefore, it is possible that the jets will be stable and propagate initially with a constant speed. 
 
For given mass accretion rate and radius, $\sigma_0$ increases with the BH spin parameter; i.e., the Poynting flux in the jets increases as the BH  rotates faster. For $\dot{m} = 1$ (Fig. \ref{sigma99}), the maximum value of $\sigma_0$ is $0.993$, which is obtained at the innermost stable orbit for a BH spin parameter of $0.9982$. (However, $\sigma_0$ increases for higher spins but we limit here the value of the BH spin parameter to that of Thorne's model.) For $\dot{m} = 10^{-1.8}$, which delimits the accretion power regime from the spin-down power regime, the initial magnetization parameter is larger than $\sim 5$ for any $a_* > 0.95$.

One can consider a relation between the magnetization parameter (Eq. \ref{sigma}) and the initial Lorentz factor of the jets of the form $\gamma_{0} =\sigma_0^{1/q}$, where the value of the power low index depends on the magnetic field configuration. For a radial outflow with negligible gas pressure, $q = 3$ \citep{michel69}. For a collimated MHD jet, the value of $q$ is also 3 if the flux distribution is the same \citep{fendt01}. With the power low above, the jets can present a radial distribution of the initial Lorentz factor, which increases with decreasing radius. 
Note that $\gamma_{0}$ differs from the bulk Lorentz factor of the jets ($\gamma$) that is defined below (Eq. \ref{eq:lF}), where it is assumed that the Poynting flux in the jets has been fully converted into kinetic energy flux. 

\begin{figure}\centering
\epsfig{file= 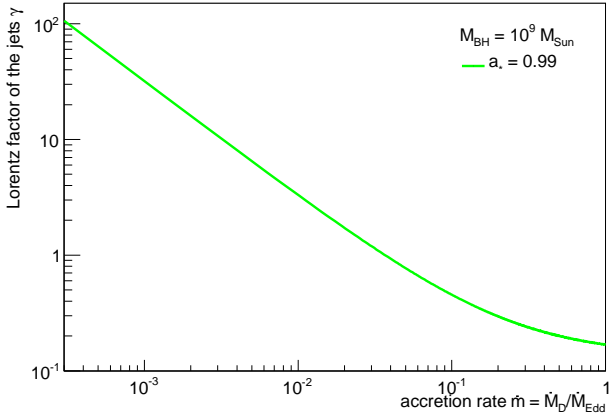,height=5.5cm}
\caption{Lorentz factor of the jets as a function of  $\dot{m}$ (Eq. \ref{eq:lF}). The jets have a bulk Lorentz factor $\gamma > 2$ when the mass accretion rate $\dot{m} < 10^{-1.8}$,  which corresponds to the spin-down power regime. In the case of the accretion power regime, $\dot{m} > 10^{-1.8}$, the jets are mildly- and sub-relativistic, $\gamma < 2$.}
\label{lFmdot}
\end{figure}

The bulk Lorentz factor of the jets $\gamma$, defined by
\begin{equation}
P_{\mathrm{jets}} = \gamma \dot{M}_{\mathrm{jets}} c^2 =  \gamma q_{\mathrm{jets}}\dot{m}\dot{M}_{\mathrm{Edd}} c^2,
\label{eq:lF}
\end{equation} 
[which follows, e.g., from \citet{fb95}; see also \citet{vila10}] is drawn in Fig. \ref{lFmdot} as a function of the mass accretion rate. The jets have a relativistic speed of $0.9-0.995\, c$ (or $\gamma = 2-10$, which is the typical bulk Lorentz factor for an AGN jet) when the mass accretion rate $\dot{m} \in [10^{-2.5}, 10^{-1.8}]$; i.e., these jets correspond to the spin-down power regime. In the case of the accretion power regime, the jets are mildly- and sub-relativistic ($\gamma < 2$). There is no significant variation of $\gamma$ with the BH spin parameter ($a_* \geqslant 0.95$) for a given mass accretion rate.

\section{Rate of the disc angular momentum removed by the jets }
\label{sec:angmom}

Now, we define the rate of the disc angular momentum removed by the jets as
\begin{equation}
J_{\mathrm{jets}}=2 \int_{r_{\mathrm{ms}}}^{r_{\mathrm{sl}}}2\pi JL^{\dagger}rdr.
\end{equation} 
Using the angular momentum conservation law (Eq. \ref{eq:angmom}), the rate of the disc angular momentum removed by the jets can be written as
\begin{equation}
J_{\mathrm{jets}}=\left( 1-q_{\mathrm{jets}}\right) {\dot{M}}_{\mathrm{D}}c\left( L^{\dagger}_{\mathrm{sl}} -L^{\dagger}_{\mathrm{ms}} \right)+4\pi \int_{r_{\mathrm{ms}}}^{r_{\mathrm{sl}}}rHdr,
\label{eq:angmom2}
\end{equation}
where the dimensionless specific angular momentum of the gas particle orbiting in the accretion disc is 
\begin{equation}
L^{\dagger}(r_*)=r_*^{-1/2}\frac{1-2a_*r_*^{-3/2}+a_*^2r_*^{-2}}{\left( 1-3r_*^{-1}+2a_*r_*^{-3/2}\right) ^{1/2}}.
\label{eq:Ldagger}
\end{equation} 

\begin{figure}
\centering
\epsfig{file= 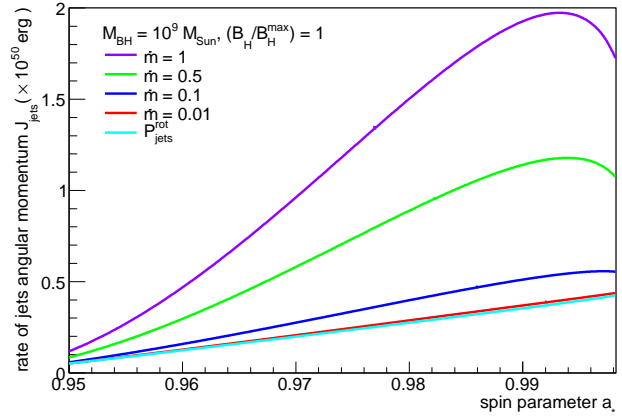,height=5.5cm}
\caption{Rate of the disc angular momentum removed by the jets (Eq. \ref{eq:Jjets}) as a function of the BH spin parameter $a_*$ for four values of the mass accretion rate. The mass accretion rates from top to bottom are 1, 0.5, 0.1 and 0.01. The bottom curve represents the BH spin-down transferred to the disc inside the ergosphere by magnetic connection. Note that there is a slight difference between the two bottom curves (red and turquoise curves).}
\label{fig:Jjets}
\end{figure}

Combining Eq. (\ref{eq:angmom2}) with (\ref{eq:omegaH}) -- (\ref{eq:magn}), (\ref{eq:cont}) and (\ref{dRdr}), the rate of the disc angular momentum removed by the jets becomes
\begin{equation}
\begin{split}
J_{\mathrm{jets}}&=\dot{m}{\dot{M}}_{\mathrm{Edd}}^{\dagger}c\,r_{\mathrm{g}}\varepsilon^{-1}(1-q_{\mathrm{jets}})\left( L^{\dagger}_{\mathrm{sl}_*}-L^{\dagger}_{\mathrm{ms}_*}\right)\left(\frac{M}{10^9M_{\odot}}\right)  \\  &+{\dot{M}}_{\mathrm{Edd}}^{\dagger}c\,r_{\mathrm{g}} \mathrm{C}_*\left( \frac{B_{\mathrm{H}}}{B_{\mathrm{H}}^{\mathrm{max}}}\right)^2\left(\frac{M}{10^9M_{\odot}}\right) \\ &\int_{r_{\mathrm{ms}_*}}^{r_{\mathrm{sl}_*}}r_*^{1-n}\mathrm{R}_*^{1/2}\left( \Omega_{\mathrm{H}_*}-\Omega_{\mathrm{D}_*}\right)dr_*,
\end{split}
\label{eq:Jjets}
\end{equation} 
where $C_*$ and $R_*$ are defined by Eq. (\ref{eq:constante}). $L^{\dagger}_{\mathrm{sl}_*}$ and $L^{\dagger}_{\mathrm{ms}_*}$ are the specific angular momentum of the gas particle (Eq. \ref{eq:Ldagger}) evaluated at the stationary limit surface and at the innermost stable orbit, respectively. We can also write the disc angular momentum removed by the jets as the sum of two components, the accretion and the rotation parts,
\begin{equation}
J_{\mathrm{jets}}=J_{\mathrm{jets}}^{\mathrm{acc}}+J_{\mathrm{jets}}^{\mathrm{rot}}.
\end{equation} 

Figure \ref{fig:Jjets} shows the rate of the disc angular momentum removed by the jets as a function of the spin parameter of the BH, given four values of the mass accretion rate ($\dot{m}=$ 1, 0.5, 0.1 and 0.01), as well as the BH spin-down $J_{\mathrm{jets}}^{\mathrm{rot}}$ (bottom curve). To know how this angular momentum is transported by the jets, further models must be employed.


\begin{figure}\centering
\epsfig{file= 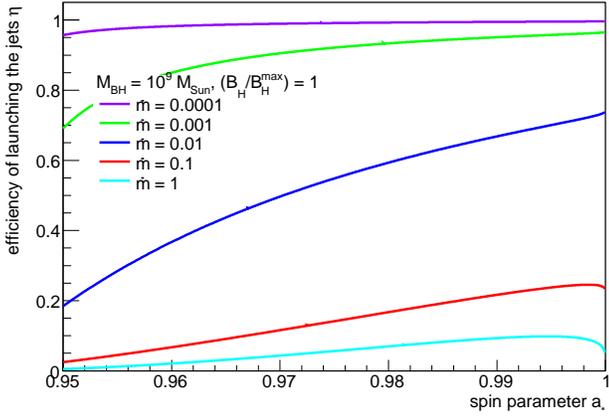,height=5.5cm}
\caption{Efficiency of jet launching (Eq. \ref{eq:effic}) as a function of the BH spin parameter $a_*$ for different mass accretion rates. The mass accretion rates from the top to bottom are 0.0001, 0.001, 0.01, 0.01 and 1. For very low mass accretion rates, $\dot{m} < 0.001$, the BH spin-down becomes a very efficient mechanism of launching the jets via the accretion disc.}
\label{fig:eta}
\end{figure}

\section{Efficiency of jet launching}
\label{sec:efficiency}

We define the efficiency of jet launching as the ratio of the launching power of the jets to the total power that comes from the gravitational energy of the accretion disc and from the rotational energy of the BH. Thus,
\begin{equation}
\eta =\frac{P_{\mathrm{jets}}}{\dot{m}{\dot{M}}_{\mathrm{Edd}}c^2 + P_{\mathrm{jets}}^{\mathrm{rot}}}.
\label{eq:effic}
\end{equation}

In Fig. \ref{fig:eta}, we plot the efficiency of jet launching for the range of the mass accretion rate $\dot{m} \in [0.0001, 1]$. For very low mass accretion rates, $\dot{m} < 0.001$, the efficiency of jet launching reaches values close to unity, in which case the spin-down of the BH becomes a very efficient mechanism to launch the jets via the accretion disc. For the spin-down power regime, the efficiency of jet launching is higher than the maximal efficiency of converting the gravitational energy of the accretion disc into radiation ($\varepsilon = 0.42$), as a result of transferring the BH rotational energy to the accretion disc via BH-disc magnetic connection.

\section{Spin evolution of the black hole}
\label{sec:evolution}

Theoretically, a Kerr BH can be spun up to a state with a spin parameter whose maximum value is $a_*=1$. As the spin evolves, a Kerr BH can achieve a stationary state. A theorem established by \citet{hawking72} states that a BH is in a stationary state if and only if the BH is either static or axisymmetric. Suppose we have a Kerr BH. Perturbing fields can, however, deflect the spin orientation away from the symmetry axis. In this case, the BH must either spin down until a static (Schwarzschild) BH is reached or evolve in such a way that it aligns its spin with the perturbative field orientation. 

Next, we study the BH spin evolution and seek the maximum spin parameter that corresponds to a stationary state of the BH, when both the BH-disc magnetic connection and the jet formation are considered. \citet{t74} calculated the influence of photon capture on the spin evolution of the BH and found a limiting state of $a_{*,\mathrm{lim}}\simeq 0.9982$. This limit does not apply to the model proposed here since the disc inside the BH ergosphere is not radiant, as in the case of Thorne's model. Instead, it drives the jets. We consider this limit only to determine the maximum value of the BH magnetic field, given at the time when the BH accretes at near the Eddington limit. 

\cite{bardeen70} showed that the mass and angular momentum of the BH can be changed by the specific energy and angular momentum of the particles carried into the BH. The BH mass (and the angular momentum) variation equals the value of the particle specific energy (and angular momentum) at the innermost stable orbit multiplied by the rest mass accreted $(dM_0)$ if no other stress energy is allowed to cross the horizon. That is,  
\begin{equation}
dM=E^{\dagger}_{\mathrm{ms}}\,dM_0\:\: \mathrm{and} \:\: dJ=J^{\dagger}_{\mathrm{ms}}\,dM_0,
\label{eq:mj1}
\end{equation} 
where $E^{\dagger}_{\mathrm{ms}}$ and $J^{\dagger}_{\mathrm{ms}}$ are the specific energy and angular momentum of the particles evaluated at the innermost stable orbit. Using Eq. (\ref{eq:mj1}), one can obtain the differential equation that describes the spin evolution of the BH due to matter accretion:
\begin{equation}
\left( \frac{da_*}{d\,\ln M}\right)_{\mathrm{matter}} =\frac{c}{GM}\left(\frac{dJ}{dM}\right)-2a_*.
\label{eq:spin1}
\end{equation}

Now, we consider the magnetic extraction of the BH rotational energy through the BH-disc magnetic connection. The spin evolution law (Eq. \ref{eq:spin1}) will be changed due to the counter-acting torque exerted on the BH by the magnetic field that connects the BH to the disc inside the ergosphere. The energy and angular momentum lost (or gained, depending on the angular velocities of the BH and disc, cf. Eq. \ref{eq:li}) by the BH through the BH-disc magnetic connection are \citep[e.g.,][]{li02}:
\begin{equation}
c^2\left( \frac{dM}{dt}\right) _{\mathrm{HD}}= 2P_{\mathrm{HD}}\,\ \mathrm{and}\ \left( \frac{dJ}{dt}\right) _{\mathrm{HD}}= 2T_{\mathrm{HD}},
\label{eq:mj2}
\end{equation} 
where $P_{\mathrm{HD}}=\Omega_{\mathrm{H}}T_{\mathrm{HD}}$, and the factor `2' comes from the fact that the accretion disc has two surfaces. Adding the effects of the BH spin-up by accretion (Eq. \ref{eq:mj1}) and the BH spin-down by magnetic connection (Eq. \ref{eq:mj2}), the equations for evolution of the BH mass and the BH angular momentum become:
\begin{equation}
c^2\left(\frac{dM}{dt}\right)=\left( 1-q_{\mathrm{jets}}\right) \dot{M}_{\mathrm{D}}c^2E_{\mathrm{ms}}^{\dagger}+c^2\left(\frac{dM}{dt}\right)_{\mathrm{HD}},
\label{eq:dMdt}
\end{equation}
\begin{equation}
\left(\frac{dJ}{dt}\right)=\left( 1-q_{\mathrm{jets}}\right) \dot{M}_{\mathrm{D}}L_{\mathrm{ms}}^{\dagger}+\left(\frac{dJ}{dt}\right)_{\mathrm{HD}}.
\label{eq:dJdt}
\end{equation}
Using these two equations, we can express the BH spin evolution as
\begin{equation}
\left( \frac{da_*}{d\,\ln M}\right)_{\mathrm{total}} =\frac{c}{GM}\frac{(1-q_{\mathrm{jets}}){\dot{M}}_{\mathrm{D}}L_{\mathrm{ms}}^{\dagger}+\left(\frac{dJ}{dt}\right)_{\mathrm{HD}}}{(1-q_{\mathrm{jets}}){\dot{M}}_{\mathrm{D}}E_{\mathrm{ms}}^{\dagger}+\left(\frac{dM}{dt}\right)_{\mathrm{HD}}}-2a_*,
\label{totspin}
\end{equation}
when both the BH spin-up by accreting matter and the BH spin-down due to the angular momentum transferred from the BH to the disc inside the ergosphere are considered. From Eqs. (\ref{eq:spin1}) and (\ref{totspin}), the spin-down of the BH by means of BH-disc magnetic connection is described by
\begin{equation}
\left( \frac{da_*}{d\,\ln M}\right)_{\mathrm{HD}}=\left( \frac{da_*}{d\,\ln M}\right)_{\mathrm{total}}-\left( \frac{da_*}{d\,\ln M}\right)_{\mathrm{matter}}.
\label{eq:spin2}
\end{equation} 

\begin{figure}\centering
\epsfig{file= 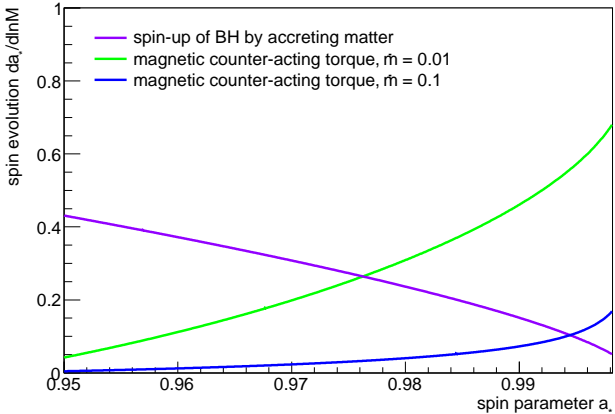,height=5.5cm}
\caption{Spin evolution of a Kerr BH. The purple line represents the driving torque by which the matter spins up the BH (Eq. \ref{eq:spin1}). The blue and green lines represent the counter-acting torque on the BH due to transfer of rotational energy from the BH to the disc (Eq. \ref{eq:spin2}) for $\dot{m}=0.1$ and $\dot{m}=0.01$, respectively. For these two cases of the mass accretion rate, the stationary state of the BH corresponds to $a_*\simeq 0.9944$ and $a_*\simeq 0.9762$, respectively.}
\label{fig:spin}
\end{figure}

Figure \ref{fig:spin} shows the spin evolution of a Kerr BH for two values of the mass accretion rate $\dot{m}=0.1$ and $\dot{m}=0.01$, respectively. The purple line represents the driving torque by which the matter spins up the BH [Eq. \ref{eq:spin1}; see also fig. 6 in \citet{t74}], and the blue and green lines represent the counter-acting torque on the BH due to transfer of rotational energy from the BH to the disc (Eq. \ref{eq:spin2}) for $\dot{m}=0.1$ and $\dot{m}=0.01$, respectively. The crossing point of the plots corresponds to the spin parameter for which the BH is in a stationary state. For $\dot{m}=0.1$, if the BH initially rotates with $a_* = 0.9982$, the BH may spin down to a stationary state with a maximum spin parameter of $a_*=0.9944$. For $\dot{m}=0.01$, the maximum spin parameter is $a_*= 0.9762$, whereas for $\dot{m}=0.001$, the maximum spin parameter is $a_*=0.9525$. Thus, as the mass accretion rate decreases, the maximum spin parameter corresponding to a stationary BH decreases as well. On the other hand, as the mass accretion rate decreases, the magnetic torque reaches values close to unity, so it is greater than 0.43, which is the maximum value of the matter torque. This implies deviations from pure Keplerian orbits, and so the possibility to drive away the excess angular momentum of the disc in the form of jets when the BH spin-down power is considered. A further analysis of the spin evolution (which is not explicitly shown in Fig. \ref{fig:spin}) suggests that a BH needs a mass accretion rate of at least $\dot{m} \sim 0.001$ for its spin to stay high ($a_* \geqslant 0.95$). For lower mass accretion rates ($\dot{m} < 0.001$), the BH may spin down continuously until the BH reaches a static state. It can spin-up again to $a_* \geqslant 0.95$ if a large amount of matter is provided by accretion (or by merging, which is not discussed here). In this case, the amount of accreted mass should be a factor of about 1.84 from the initial mass of the BH \citep{t74}.

\section{Maximum lifetime of the AGN from the BH spin-down power}
\label{sec:relevance}

In this section, we calculate the time-scale needed for a Kerr BH to spin down from $a_* \sim 1$ to 0.95, which can then be related to the maximum lifetime of the AGN, provided that the BH was spun up to nearly its maximum spin during a phase when the AGN was active. The AGN can be active as long as the gravitational energy of the accretion disc is converted into observed radiation energy. Such an AGN can have a longer lifetime through the additional use of its BH spin-down power, despite having a very-low mass-accretion rate ($\dot{m} < 0.1$). Following the well-known work by \citet{salpeter}, the time needed to fuel the AGN to a bolometric luminosity $L_{\mathrm{bol}} \sim 10^{45}$ erg s$^{-1}$ can be $\sim 10^{7}$ yr for a typical radiative efficiency of $\varepsilon=0.1$. Moreover, the lifetime of high accreting AGN (and quasars) was constrained by recent observations to the range $ \sim 10^{7} - 10^{8}$ yr \citep[e.g.,][]{porciani,hopkins09}.

Next, we estimate the maximum lifetime of the AGN. Differentiating the BH angular momentum $J = M c a = (GM^2/c)\,a_*$ with respect to time $t$, the BH time evolution is specified by
\begin{equation}
\left( \frac{da_*}{dt}\right) = \frac{c}{GM^2}\left( \frac{dJ}{dt}\right)-2\frac{a_*}{M} \left( \frac{dM}{dt}\right).
\label{eq:tevol}
\end{equation} 
Integrating this equation, the time interval over which the BH spin evolves between two given values of $a_*$ is
\begin{equation}
t = \int_{{a_*}_1 }^{{a_*}_2}\left[ \frac{c}{GM^2} \left( \frac{dJ}{dt}\right)-2\frac{a_*}{M}\left( \frac{dM}{dt}\right)  \right]^{-1} da_*,
\label{time}
\end{equation} 
where $(dJ/dt)$ can be obtained from Eqs. (\ref{eq:dJdt}) and (\ref{eq:spin2}), and $(dM/dt)$ from Eqs. (\ref{eq:dMdt}) and (\ref{eq:spin2}). With the above equation, we can estimate the lifetime of the AGN. The time interval is not dependent on the BH mass.

\begin{figure}\centering
\epsfig{file= 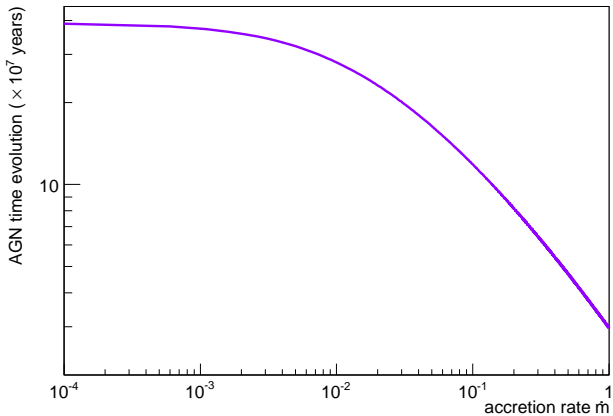,height=5.5cm}
\caption{Lifetime of the AGN from the BH spin-down power as a function of the mass accretion rate. The lifetime of the AGN is $\sim 3\times10^{7}$ yr when using the accretion power. The BH spin-down power adds to the lifetime of the AGN, for instance, $2.8\times10^{8}$ yr, $3.9\times10^{8}$ yr and $\sim 4\times10^{8}$ yr when $\dot{m} = 10^{-2}$, $\dot{m} = 10^{-3}$ and $\dot{m} < 10^{-4}$, respectively (while the BH spin decreases from $a_* \sim 1$ to 0.95).}
\label{evolution}
\end{figure}

In Fig. \ref{evolution}, we plot the time evolution of the AGN as a function of the mass accretion rate (Eq. \ref{time}), when the BH spin parameter decreases from $a_* \sim 1$ to 0.95. For a mass accretion rate close to the Eddington limit, the lifetime of the AGN is about $3\times10^{7}$ yr. The lifetime curve moves toward lower mass accretion rates for another $\sim 10^{8}$ yr, when the AGN uses its BH spin-down power to launch the jets. Therefore, the total lifetime of the AGN can be much longer than the Hubble time ($t_{\mathrm{H}} \sim 10^{10.14}$ yr). The maximum lifetime of the AGN is, however, dependent on the mass accretion rate. The maximum lifetime of the AGN from the BH spin-down power is, for instance, $\sim 2.8\times10^{8}$ yr, $\sim 3.9\times10^{8}$ yr and $\sim 4\times10^{8}$ yr for $\dot{m} = 10^{-2}$, $\dot{m} = 10^{-3}$ and $\dot{m} < 10^{-4}$, respectively. In the latter case, the BH may not attain a stationary state and spins down until a static BH is reached. The lifetime of the AGN scales with the strength of the BH magnetic field relative to its maximum value as $ t \sim (B_\mathrm{H}/B_\mathrm{H}^{\mathrm{max}})^{-2}$. Therefore, if the $B_\mathrm{H}$ is a factor of $k$ lower than $B_\mathrm{H}^{\mathrm{max}}$, the maximum lifetime of the AGN will be a factor of $k^2$ larger. For instance, when $k = 7$ and $\dot{m} = 10^{-2}$, one obtains the exact Hubble time.

Now, we compare our results to the lifetime of an AGN powered by the Blandford--Znajek mechanism. The total energy that can be extracted by the Blandford--Znajek mechanism is \cite[e.g.,][]{li00a}:
\begin{equation}
E_{\mathrm{BZ}} \simeq 0.09 Mc^2 \simeq 1.6 \times 10^{62} \; \mathrm{erg}\;\left( \frac{M}{10^9 M_{\odot}}\right).
\end{equation} 
The maximum rate of energy extraction by the Blandford--Znajek mechanism is [Eq. 4.50 of \citet{membrane}]:
\begin{equation}
P_{\mathrm{BZ}} \simeq 10^{45} \; \mathrm{erg} \mathrm{s}^{-1}\;\left( \frac{B_{\mathrm{H}}}{10^4 \mathrm{G}}\right)^2 \left( \frac{M}{10^9 M_{\odot}}\right)^2,
\end{equation} 
for a BH with $a_* \sim 1$. Therefore, the lifetime of an AGN powered by the Blandford--Znajek mechanism is:
\begin{equation}
t_{\mathrm{BZ}} = \frac{E_{\mathrm{BZ}}}{P_{\mathrm{BZ}}} \simeq 5\times 10^9 \; \mathrm{yr} \left( \frac{B_{\mathrm{H}}}{10^4 \mathrm{G}}\right)^{-2} \left( \frac{M}{10^9 M_{\odot}}\right)^{-1},
\end{equation} 
which is not dependent of the BH mass, as $B_{\mathrm{H}}$ scales with $( M/10^9 M_{\odot} )^{-1/2}$. Except for the exact time-scale, our result (Eq. \ref{time}) scales with the magnetic field and with the BH mass exactly the same way as for the Blandford-Znajek mechanism.

Moving back to our results, if the mass accretion rate changes over the whole life of the AGN from $\dot{m} \sim10^{-1.8}$ to $ \dot{m} < 10^{-4}$, the maximum lifetime of the AGN can be even longer than that of an AGN powered by the Blandford--Znajek mechanism. In summary, the maximum lifetime of the AGN can be much longer than $\sim 10^{7}$ yr when using the BH spin-down power. The lifetime is dependent on the mass accretion rate, as well as on the factor $(B_\mathrm{H}/B_\mathrm{H}^{\mathrm{max}})$. We mention that the results presented here refer only to rapidly-spinning BHs ($a_* \geqslant 0.95$).

\section{Summary and conclusions}
\label{sec:summary}

Starting from the general-relativistic conservation laws for matter in a thin accretion disc, we included both the BH-disc magnetic connection and the jet formation. The jets are launched from the disc inside the BH ergosphere, where the rotational effects of the space time become much stronger. In the BH ergosphere, the frame-dragging effect ensures that the magnetic field lines frozen in the disc co-rotate everywhere with the BH, reducing the effect of magnetic reconnection and keeping the magnetic field configuration globally the same. Furthermore, the jets can extract mass, energy and angular momentum from the inside the ergosphere disc. For this situation, we derived the mass flow rate into the jets, the launching power of the jets, the angular momentum removed by the jets, the efficiency of launching the jets, the maximum spin parameter attained by a stationary BH and the maximum lifetime of an AGN. 

\begin{itemize}
\item We found that the mass flow rate into the jets is dependent on the BH spin parameter, where the mass outflow into the jets can be associated with the rotation of the space-time itself. For near maximal spin parameter $a_* \sim 1$, the mass flow rate into the jets is about 98 per cent of the available mass flow in the disc inside the BH ergosphere, whereas for $a_* = 0.95$ this is only about 8 percent. This means that in the case of maximal spin, the BH almost stops being fed by accreting matter. As a possible alternative, the jets may have no matter right at the beginning (i.e., they are Poynting flux jets) and get it only very quickly from drifts just above the disc or the surrounding wind (i.e., indirectly from the disc). This may only be relevant for extremely low accretion rates.

\item In this work, we considered the case of rapidly-spinning BHs with a spin parameter of $a_* \geqslant 0.95$ and a mass of $10^{9} M_{\odot}$, and we assumed that the power of the disc inside the BH ergosphere is used to drive the jet. We found that at low mass accretion rates, the jet power can be supplied by the BH rotational energy via the disc inside the ergosphere. The switch from an accretion power regime to a spin-down power regime corresponds to a mass accretion rate of $\dot{m} \simeq 10^{-1.8}$. In the case of the spin-down power regime ($\dot{m} < 10^{-1.8}$), the jet power is $\sim 10^{45}$ erg s$^{-1}$, which is only $10^{-2}$ of the Eddington luminosity of a $10^{9}$ solar mass BH. This is comparable to the maximum rate of energy extraction by the Blandford--Znajek mechanism, which is $\sim 10^{45}$ erg s$^{-1}$ for a BH mass of $10^{9} M_{\odot}$ and $a_* \sim 1$. This implies that, in principle, both the Blandford--Znajek mechanism and launching jets from the disc inside the ergosphere via the BH-disc magnetic connection can operate. We intend to study the driving of the jets when the BH is threaded by a combination of open and closed magnetic field lines in future work.

\item The jets can have a relativistic speed, $0.9-0.995\, c$ (or $\gamma = 2-10$, which is the typical bulk Lorentz factor for an AGN jet), when the mass accretion rate $\dot{m} \in [10^{-2.5}, 10^{-1.8}]$. In the case of the accretion power regime, the jets are mildly- and sub-relativistic. However, after launching the jets can be accelerated through magnetic processes. The jets remove the angular momentum of the disc inside the BH ergosphere at a rate which is dependent on the BH spin parameter. To know how this angular momentum is transported by the jets, further models have to be employed. The efficiency of jet launching is higher at low mass accretion rates, reaching values close to unity for $ \dot{m} \sim 10^{-4}$. In this case, the BH spin-down power is efficiently used to launch the jets.

\item Considering the balance between the BH spin-up by accreting matter and the BH spin-down due to the magnetic counter-acting torque on the BH, we determined the maximum spin parameter which corresponds to a BH stationary state. The maximum spin value shifts towards $a_* = 0.95$ as the mass accretion rate decreases. For instance, the maximum spin parameter corresponding to $\dot{m}=0.1$, $\dot{m}=0.01$ and $\dot{m}=0.001$ is $a_*=0.9944$, $a_*= 0.9762$ and $a_*=0.9525$, respectively. At lower mass accretion rates ($\dot{m} < 0.001$), the BH may undergo a spin-down process towards a static BH. The BH never reaches a stationary state unless a large amount of matter is provided (perhaps by star capture or by merging) to spin up the BH again to $a_* \geqslant 0.95$. 

\item We showed that an AGN can have a much longer lifetime than $\sim 10^{7}$ yr when using the BH spin-down power, and the maximum lifetime is dependent on the mass accretion rate, as well as on the factor $(B_\mathrm{H}/B_\mathrm{H}^{\mathrm{max}})$. After an accretion-dominated phase of about $3\times10^{7}$ yr, the AGN can live off of the BH spin-down power for another $10^{8}$ yr. The BH spin-down power adds to the lifetime of the AGN, for instance, $\sim 2.8\times10^{8}$ yr, $\sim 3.9\times10^{8}$ yr and $\sim 4\times10^{8}$ yr for $\dot{m} = 10^{-2}$, $\dot{m} = 10^{-3}$ and $\dot{m} < 10^{-4}$, respectively. Moreover, if the $B_\mathrm{H}$ is a factor of $k$ lower that $B_\mathrm{H}^{\mathrm{max}}$, then the lifetime of the AGN will be a factor of $k^2$ larger. For $k = 7$ and $\dot{m} = 10^{-2}$, one obtains the exact Hubble time. Another possibility is that the mass accretion rate changes over the whole life of the AGN from $\dot{m} \sim10^{-1.8}$ to $ \dot{m} < 10^{-4}$. In this case, the maximum lifetime of the AGN can be even longer than that of an AGN powered by the Blandford--Znajek mechanism, which is $\sim 5\times 10^9$ yr for a BH with $a_*$ close to the maximal value. However, it will be difficult to predict a maximum lifetime of the AGN for this possibility, since there is no mechanism, to date, to control the change of the mass accretion rate over long intervals of time.
\end{itemize}

In the limit of the BH spin-down regime, the model proposed here can be regarded as a variant of the Blandford-Znajek mechanism. It is a variant of the Blandford-Znajek mechanism if an insignificant fraction of the rotational energy is transported away by jets through open magnetic field lines that thread the BH. If we were to consider the open magnetic field lines in addition to the closed magnetic field lines that thread the BH, our calculations would have been completed such that they would include the Blandford-Znajek mechanism.

The results presented in this paper are dependent on our assumptions that a BH-disc magnetic connection exists. Closed magnetic field lines in the BH ergosphere may be produced by a current ring in the vicinity of the BH.  Models for the magnetic connection where a poloidal magnetic field is generated by a single electric current flowing in the BH equatorial plane or at the inner edge of the accretion disc were proposed, for instance, by \citet{li02R} and \citet{wang}. The key parameters of the proposed model are, however, the BH mass, the BH spin and the mass accretion rate. 

One indirect way to test whether this mechanism operates in reality is to study the relation between the observed radio flux density from AGN (e.g., from flat-spectrum core sources) and their mass accretion rates in order to fit the model prediction with respect to the relation between the power of the jet and the mass accretion rate (see Fig. \ref{PjetsVsMdot}). That relation shows that the power of the jet does not depend linearly on the mass accretion rate all the way down to very low accretion rates, so that there can be sources with relatively strong jet power but low mass accretion rate. In this case the jet power is mainly dependent on the BH parameters, such as the mass and the spin of the BH. The first step is to find a relation between the power of the jets and the observed radio flux density, $F_{\rm obs}$. For a conical jet (where the magnetic field along the jet scales with the distance, $B \sim z^{-1}$, and the electron number density in the jet scales as $C' \sim z^{-2}$) from a flat-spectrum core source, we found the dependence:  $P_{\rm jets} \sim F_{\rm obs}^{6/5} D^{12/5} M^{-7/10}$, where $D$ is the distance to the AGN and $M$ is the BH mass. A full description of its derivation is the subject to a paper in preparation. The second step is to produce a complete sample of AGN with known jet parameters, as the Doppler factor, and whose mass accretion rate can be constrained by observational data, and then to fit the model prediction. 

Furthermore, if AGN were in the Poynting flux limit disregarding the mass accretion rate, then one might expect that the power in the jet from a large sample runs into a lower limit, which is given by the minimum Poynting flux.

The model presented here can also be extended to microquasars assuming that physical quantities (e.g., BH magnetic field, jet power, etc.) scale with the BH mass. 

Although numerical simulations of jet formation from the ergosphere of a rapidly-spinning BH with closed magnetic field lines that connect the BH to the disc inside the ergosphere has not been performed yet, this can be one of the challenges to be faced by numerical relativists.

\section*{Acknowledgements}
The author would like to thank Peter L. Biermann. This research was supported through a stipend from the International Max Planck Research School (IMPRS) for Astronomy and Astrophysics at the Universities of Bonn and K\"{o}ln. The author appreciates the support from MPIfR during the last phase of this work.

\bibliographystyle{mn2e}
\bibliography{MCmodel.bib}

\appendix
\section{Derivation of equation 4}
\label{conslaws}

Here, we present the derivation of the angular momentum conservation law for the matter that flows in the accretion disc (Eq. \ref{eq:angmom}). This derivation is based on the general-relativistic angular-momentum-conservation law that describes the structure of a geometrically thin accretion disc \citep{pt}. Based on this conservation law, \citet{li02} derived the conservation law that includes the BH-disc magnetic connection, and \citet{db} derived the conservation law that includes the jet. In a slightly different manner, our result is obtained when both the BH-disc magnetic connection and the jet formation are considered. 

The angular-momentum conservation law of the matter in a standard, thin accretion disc is given by eq. (23) of \citet{pt}. For details on this procedure and the meaning of the notations used, the reader is referred to the paper by  \citet{pt}. Below, we consider the first and second integrals of their equation, keeping the non-vanishing terms:
\begin{equation}
\begin{split}
& \left\lbrace\int_{-h}^{h}\limits \left( 2\pi \Delta t\right) \left( <\rho_0>u_{\phi}u^r + <t^r_{\phi}> \right) \sqrt{|g|} dz\right\rbrace_{r}^{r+\Delta r}\\
& =   \left\lbrace ( 2\pi \,\Delta t) ( \Sigma  L^{\dagger} u^r + W_{\phi}^{r}) \sqrt{|g|} \right\rbrace_{r}^{r+\Delta r}\\
& = \Delta t \,\Delta r\left[ -( \dot{M}_{\mathrm{D}} - \dot{M}_{\mathrm{jets}})  L^{\dagger} + 2\pi r W_{\phi}^{r} \right]_{,r},
\end{split}
\end{equation} 
where we made use of the rest-mass conservation law that includes the mass flow into the jets , $\dot{M}_{\mathrm{D}} - \dot{M}_{\mathrm{jets}} = -2\pi r \Sigma u^r$, and 
\begin{equation}
\begin{split}
& \left\lbrace\int_{r}^{r+\Delta r}\limits \left( 2\pi \Delta t\right) \left( <\rho_0> u_{\phi} u^z+u_{\phi} <q^z> \right)  \sqrt{|g|} dr\right\rbrace_{-h}^h\\
& = 2\,( 2\pi \,\Delta t) ( \Sigma u^z +  F )  L^{\dagger} \sqrt{|g|} \Delta r = 2\,( 2\pi r \,\Delta t \,\Delta r ) J  L^{\dagger},
\end{split} 
\end{equation} 
where $J =  \Sigma u^z +  F $ denotes the total flux of energy (of particle and electromagnetic origin) transported by jets.

Adding the two integrals, eq. (23) of Page \& Thorne becomes:
\begin{equation}
\left[ ( \dot{M}_{\mathrm{D}} - \dot{M}_{\mathrm{jets}})  L^{\dagger} - 2\pi r W_{\phi}^{r} \right]_{,r} = 4 \pi J  L^{\dagger}.
\label{23}
\end{equation}  

Now, using the definition of the magnetic torque produced by the BH on both surfaces of the accretion disc which is given by Li (see Eqs. \ref{torque} and \ref{eq:li}) and including $c$, Eq. \ref{23} becomes:
\begin{equation}
\frac{d}{dr}\left[ \left( 1-q_{\mathrm{jets}}\right) {\dot{M}}_{\mathrm{D}}cL^{\dagger}\right]+4\pi rH =4\pi rJL^{\dagger},
\end{equation} 
when both the BH-disc magnetic connection and the jet formation are considered.

\end{document}